\documentclass[fleqn,usenatbib]{mnras}

\usepackage{newtxtext,newtxmath}

\usepackage[T1]{fontenc}
\usepackage{lscape}
\usepackage{ae,aecompl}
\usepackage[title]{appendix}
\usepackage{multirow}

\usepackage{caption}
\usepackage[utf8]{inputenc}
\usepackage{graphicx}	
\usepackage{float}
\usepackage{wrapfig}
\usepackage{fancyhdr}
\usepackage{nameref}
\usepackage{amsmath}	
\usepackage{subcaption}
\usepackage{systeme}

\title[]{Observational constraints to the dynamics of dust particles in the coma of comet 67P/Churyumov-Gerasimenko}

\author[Frattin E. et al.]{Frattin E.,$^{1}$\thanks{E-mail: elisa.frattin@unipd.it}
Bertini I.,$^{2,3}$
Ivanovski S.L., $^{4}$
Marzari F., $^{1}$ 
Fulle M., $^{4}$
Zakharov V.V.,$^{2,3}$
\newauthor
Moreno F.,$^{5}$
Naletto G.,$^{1}$
Lazzarin M.,$^{1}$
Cambianica P.,$^{6}$
Cremonese G., $^{6}$
Ferrari S.,$^{7}$
\newauthor
Ferri F.,$^{7}$
G{\"u}ttler C., $^{8}$
La Forgia F., $^{1,2}$
Lucchetti A.,$^{6}$
Pajola M.,$^{6}$
Penasa L.,$^{7}$
Rotundi A., $^{2,3}$
\newauthor
Sierks H.,$^{8}$
Tubiana C. $^{3,8}$
 \\
$^{1}$ Dipartimento di Fisica e Astronomia G.Galilei, Università di Padova, Vicolo dell'Osservatorio 3, 35122 Padua, Italy.\\
$^{2}$ Dipartimento di Scienze e Tecnologie, Università di Napoli “Parthenope”, CDN, IC4 80143 Naples, Italy.\\
$^{3}$ Institute for Space Astrophysics and Planetology (IAPS)-INAF.
Via Fosso del Cavaliere 100 00133 Rome, Italy.\\
$^{4}$ Osservatorio Astronomico di Trieste,  Istituto Nazionale di Astrofisica, Via G.B. Tiepolo, 11 I-34143 Trieste, Italy.\\
$^{5}$ Instituto de Astrofísica de Andalucía, CSIC,
Granada 18008, Spain.\\
$^{6}$ Osservatorio Astronomico di Padova, Istituto Nazionale di Astrofisica, Vicolo
dell'Osservatorio 5, 35122 Padua, Italy.\\
$^{7}$ Center of Studies and Activities for Space (CISAS) G. Colombo, University of Padova, Via Venezia 15, 35131 Padua,
Italy.\\
$^{8}$ Max Planck Institute for Solar System Research, Justus-von-Liebig-Weg 3, 37077 G{\"o}ttingen, Germany.\\
}
\date{Accepted XXX. Received YYY; in original form ZZZ}
\pubyear{2019}
\begin{document}
\label{firstpage}
\pagerange{\pageref{firstpage}--\pageref{lastpage}}
  \maketitle
\begin{abstract}
In this work we aim to characterise the dust motion in the inner
coma of comet 67P/Churyumov-Gerasimenko to provide constraints for theoretical 3D coma models.
The OSIRIS camera on board the Rosetta mission was able for the first time to  acquire images of single dust particles from inside the cometary coma, very close to the nucleus.
We analyse a large number of particles, performing a significant statistic of their behaviour during the post perihelion period, when the spacecraft covered distances from the nucleus ranging between 80 and 400 km.
We describe the particle trajectories, investigating their orientation and  finding highly radial motion with respect to the nucleus.
Then, from the particle brightness profiles, we derive a particle rotational frequency of $\nu < 3.6$ Hz, revealing that they are slow rotators and do not undergo fragmentation.
We use scattering models to compare the observed spectral radiance of the particles with the simulated ones in order to estimate their size, finding values that range from millimetres up to  centimetres.
The statistics performed in this paper provide useful parameters to constrain the cometary coma dynamical models.

\end{abstract}
 

\section{Introduction}
The cometary coma is a complex environment where  several physical and chemical processes regarding solid and gaseous materials emitted from the nucleus take place, such as   fragmentation, sublimation and photoionization.  Many models of different levels of complexity have been developed during the years in order to describe the cometary dust motion   \citep{crifo2005,zakharov2009,marschall_model,stavro2017model}.
The main forces that govern the motion of the particles in the vicinity of the nucleus are the gravitational force exerted by the comet, the aerodynamical force generated by the gas drag, the solar gravity as well as the solar radiation pressure.\\
The gas is considered to be the main driver of the motion up to a certain distance from the nucleus at which the dust particles reach their terminal velocity decouple  from the gas, which occurs at about 10 nucleus radii from the center of mass \citep{crifo2002, crifo2005,zakharov2018}. \cite{FulleIvanovski2015} and \cite{stavro2017model} computed the exact distances where single particles reach their terminal velocity for 67P/Churyumov-Gerasimenko (hereafter 67P), obtaining values that range between 10 to 100 km depending on the gas field, the  mass and density of the particles. The gas production rate governs the release process and it changes along the comet trajectory before and after perihelion.\\
In general, since   drag and gravity decrease with the distance from the nucleus, at a certain height the solar pressure force finally dominates: this may happen either close or far from the nucleus, depending on the particular case \citep{moreno2017,stavro2017model,skorov2018paper2}.\\
Furthermore, the shape of the particles plays an important role in the dynamical models. 
The up-to-date observational data revealed limitations of the models which assume sphericity of dust grains. 
Indeed, the evidence  of irregular shape of dust particles was revealed by the Stardust mission, by the  IDPs (Interplanetary Dust Particles) analysis \citep{rotundi2007idp, rotundi2008stardust} and by the measurements obtained by the Grain Impact Analyzer and Dust Accumulator (GIADA, \citep{colangeli2007,DellaCorte2014}) and Cometary Secondary Ion Mass Analyzer (COSIMA, \citep{leroy2015}) for the Rosetta mission 
\citep{langevin2016,fulle_blum,stavro2017model}. The images provided by the Optical, Spectroscopic, and Infrared Remote Imaging System (OSIRIS, \citep{keller_osiris})  revealed indications of rotational motion of   particles by the brightness
variation in the tracks of single particles, interpreted as rotational motion of non-spherical particles \citep{FulleIvanovski2015,frattin}.\\
Therefore refined models, based on the use of aspherical particles, have been developed by \cite{stavro2017model} and \cite{skorov2016paper1}.\\
All the aforementioned models dealing with dust dynamics in the inner coma need to be verified, tested, and refined through observational data. It is therefore of pivotal importance to use Rosetta data to constrain theoretical interpretations providing inputs to the models.
During orbiting around comet 67P, the OSIRIS camera on board the Rosetta spacecraft \citep{keller_osiris}  was able to acquire images   from  inside the cometary coma, close to the nucleus. In  July 2015 - January 2016 time frame, the  spacecraft was at distances  between 80 km and 400 km from the nucleus (with one configuration at 1200 km).
For the first time, it was possible to directly observe tracks of single dust particles and measure their light curve, to  determine hints on their rotational state, and size, providing a good statistics of the aforementioned parameters, thanks to  the huge amount of data collected.\\
In this work we present an extensive analysis of dust particles characteristics and dynamics.
In Section \ref{geometry} we describe the dataset. In Section \ref{radial} we show the observational evidence of the radial motion of the dust particles, in Section \ref{rotational} we present a statistic of the  rotational frequency of the particles in the inner coma of the comet. Afterwards,  we  give an estimate of their shape evaluating their flattening. \\
In Section \ref{photometry} we provide a photometric analysis of the dust particles deriving their size.
A description of the rotational dynamics of the dust is given in Section \ref{section:stavro}. The conclusions are given in Section \ref{conclusion}.

\begin{figure}
    \centering
    \includegraphics[width=\linewidth]{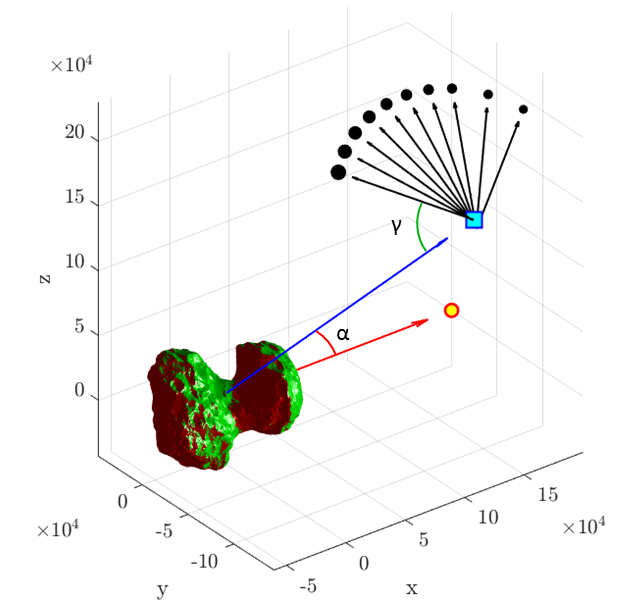}
    \caption{Example of the observational configuration of the set of images  described in Table \ref{table:radiality}, the case of STP 081 GRAIN\_COLOR\_002. The red arrow represents the direction of the Sun (yellow circle). The black arrow represents the line of sight of the camera (cyan square) for each image, while the black circles stand for the dust particles. The line of sight lies on a plane perpendicularly to the plane containing the Sun, the comet and the s/c.}
    \label{fig:geo_81_GC2}
\end{figure}

\begin{figure}
    \centering
    \includegraphics[width=\linewidth]{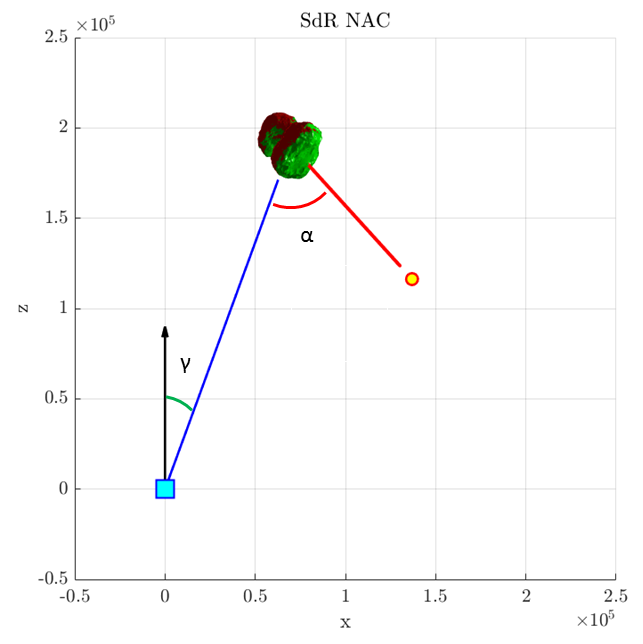}
    \caption{Example of the observational configuration of the set of images  described in Table \ref{table:periods}, the case of STP 081 GRAIN\_TRACK\_003. The red arrow represents the direction of the Sun (yellow circle). The black arrows represent the line of sight of the camera (cyan square). The line of sight lies on the plane containing the Sun, the comet and the s/c. }
    \label{fig:geo_81_GT3}
\end{figure}

 \section{Data} \label{geometry}
To study the direction of motion and the rotational state of the particles we used images acquired from August 2015 to January 2016 (the perihelion passage was in August 2015) by the OSIRIS Narrow Angle Camera (NAC) in the Orange filter ($\lambda = 649 $ nm), with an exposure time of 12.5 seconds.\\
For the analysis we used 92 radiometric calibrated and geometric distortion corrected OSIRIS Level 3 and 3B (CODMAC L4) images \footnote{The data are available at the Planetary Science Archive of the European Space Agency under https://www.cosmos.esa.int/web/psa/rosetta}.
Details about the data processing, including all calibration steps applied to the images, can be found in \citet{ceci_filtri}.
The only difference between the used image levels is solely the unit: OSIRIS Level 3 images are in  W/(m$^2$ sr nm)  while L3B images are I/F, thus unity.\\
During the entire mission, we regularly acquired
sequences of images named dust particle track and dust particle colour,which were designed to study the motion and the colours, respectively, of individual particles in the dust coma.
The design of those sequences was the same thorough the mission, allowing to combine data acquired at different epochs.\\
In this work we analyse both type of sequences. The 
activity name of each sequence is listed
in Tables \ref{table:radiality} and \ref{table:periods}.
At the time of the observations, the Rosetta spacecraft was at a distance from the nucleus between 84 and 437 km.
The Rosetta +Z-axis, which is aligned with the camera boresight direction,  was off-pointing with respect to the nucleus, which is therefore outside the camera field of view.
To study the direction of the dust motion, in Section \ref{radial}, we have analysed  28 images belonging to  5 sets of images listed in Table 1. 
The nucleus phase angle $\alpha$  (i.e. the angle between the direction of Sun and the direction of spacecraft
as seen by the nucleus) is reported here for geometrical purposes and in these images is always less than $90^{\circ}$, hence the spacecraft is located in the day side of the coma, see Figure \ref{fig:geo_81_GC2}.\\
The plane of sight is perpendicular to the plane defined by the Sun, the nucleus and the spacecraft
and these planes intersect at the line given by the direction between the Sun and the spacecraft.
Therefore, the plane of sight contains the Sun.
The images elongation $\epsilon$ (i.e. the angle between the direction of the line of sight and the direction of the Sun as seen by the  spacecraft)  changes within the sequence from the solar direction toward the anti-solar direction, from 50$^{\circ}$ to  160$^{\circ}$. 
It means that the camera points dust that scatters from the forward direction up to the backward direction with respect to the Sun. 
This allows to investigate different parts of the coma at the same epoch. 
The phase angle of the dust follows the relation $\alpha_{dust}$ = 180$^{\circ}$ - $\epsilon$.\\
To study the rotational state  and the photometry of the dust particles, in Section \ref{rotational} and \ref{photometry}, we  have analysed 64 images belonging to 7 sets of images taken in various epochs, listed in  Table \ref{table:periods}. 
We decided to focus on sets of images where the observational geometry configuration formed by the spacecraft, the comet, and the Sun  remained constant.
This means that the phase angle $\alpha$, the elongation of the Sun $\epsilon$ and the angle $\gamma$ (i.e. the angle between the direction of the spacecraft and the direction of the line of sight) are fixed within the observational series, see Figure \ref{fig:geo_81_GT3}. 
In this way, the condition of illumination of the particles  during the time interval of a series is the same.

\subsection{Detection method}

\begin{figure}
    \centering
        \includegraphics[width=\linewidth]{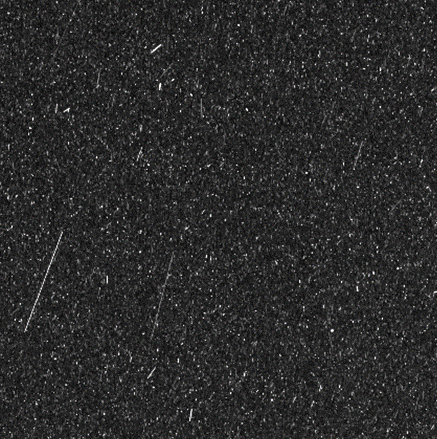}
    \caption{Image taken by NAC on 30 November 2015 at UT 19:25:04.078, belonging to STP 084. The colour scale follows a square root variation. The grey scale goes from black to white, which correspond to flux values of 0 and 0.000185  W/(m$^2$ nm sr), respectively. 
    The size of the image is 440x440 px.} 
    \label{fig:detection1}
\end{figure}

\begin{figure}
    \centering
    \includegraphics[width=\linewidth]{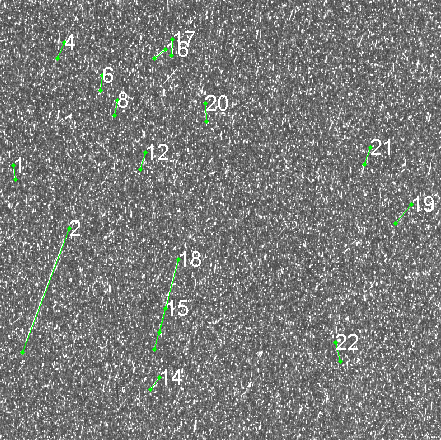}
    \caption{Last step of the detection process on image taken by NAC on 30 November 2015 at UT 19:25:04.078. The tracks are identified with a number and highlighted in green. The colour scale follows a square root variation. }
    \label{fig:detection2}
\end{figure}

The OSIRIS NAC camera is able to observe single dust particles in the coma of 67P.
Each particle draws a track on the plane of the image, due to its motion during the exposure time.
Therefore, the track is the 2D projection on the CCD focal plane of the 3D particle motion. 
We use the automatic method  developed by \cite{frattin} to detect the tracks from the images and to provide some photometric and geometric parameters, as their radiance, full width half maximum (FWHM) and  orientation.\\ 
Figure \ref{fig:detection1} shows an example of an OSIRIS image with dust tracks, while Figure \ref{fig:detection2} illustrates the final step of the detection process in which the tracks are identified with a green line and a number.
The method is based on a similarity function that evaluate the correspondence between two images.
A set of synthetic tracks is generated, characterised by their orientation and FWHM, given by the convolution of a rectangular function with a bi-dimensional Gaussian.
Then, the Normalised Cross Correlation allows to evaluate the similarity between the features identified on a local zone of the images and the synthetic tracks generated.
Eventually the length of the tracks, their position, orientation, flux and FWHM are provided.
The flux is computed subtracting the background which is evaluated considering a rectangular area around the track, with sides parallel to the sides of the image and with length and height 10 pixel longer than the projection of the track on the two dimensions, in order to include the track properly.


\section{Translational motion of the dust} \label{radial}
 In this section, we study the translational motion of the dust in the inner coma of comet 67P. We measure the particles direction of motion and we compare it with the trajectories computed by an appropriate  dynamical model, in order to describe the inner coma dust motion. 
 The data set consists in the post-perihelion images reported in  Table \ref{table:radiality}, which  geometry is described in Section \ref{geometry}.

\begin{figure}
    \centering
    \includegraphics[width = \linewidth]{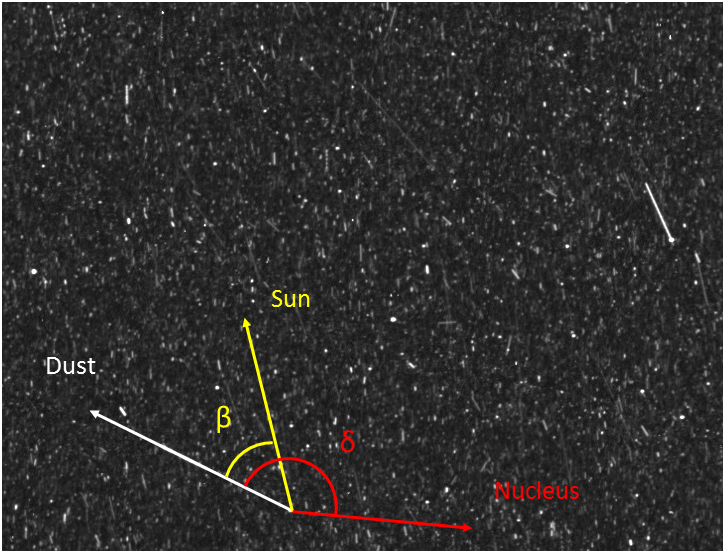}
    \caption {Single NAC image taken on 30 November 2015, STP 084, at UT 18:51:43.22. The image shows the construction of the position vectors used to compute the direction of motion of the dust. The track generated by the passage of a dust particle in the coma is shown in white,   the unit vector of the position of the Sun  in yellow and the unit vector of the position of the nucleus in red.  The colour scale follows a square root variation. The grey scale goes from black to white, which correspond to flux values of 0 and  0.000035 W/(m$^2$ nm sr), respectively. The size of the image is 722x541 px.}
    \label{fig:technique}
\end{figure}

\begin{figure*}
  \begin{subfigure}[t]{0.33\textwidth}
 \centering
 \includegraphics[width=\linewidth]{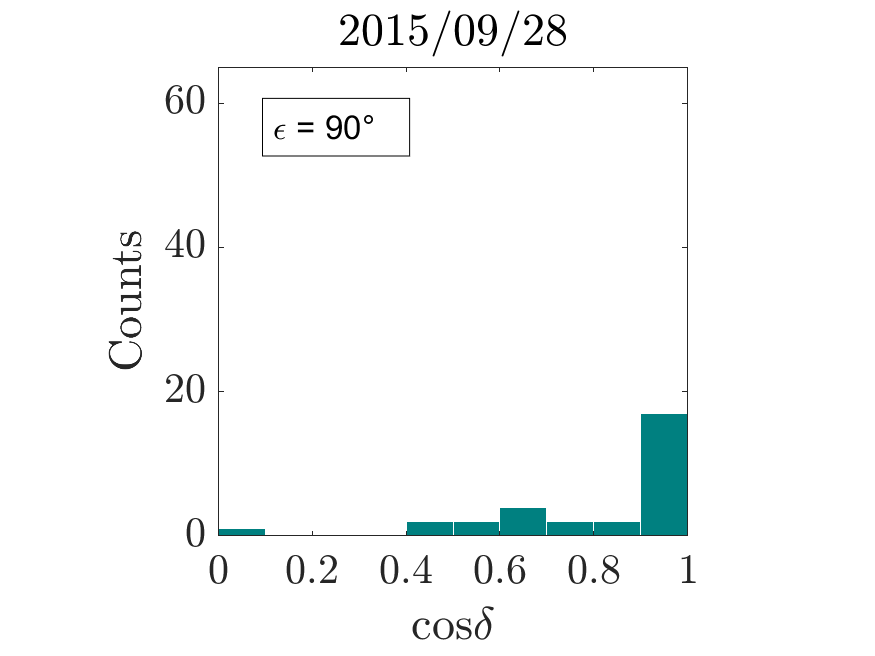}
  \end{subfigure}
 \begin{subfigure}[t]{0.33\textwidth}
 \includegraphics[width=\linewidth]{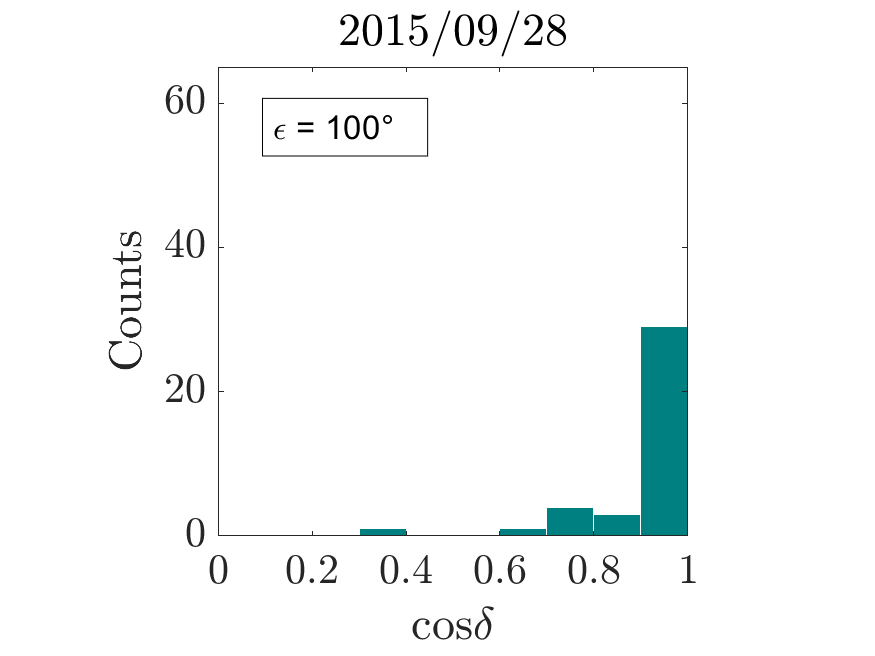}
  \end{subfigure}
  \begin{subfigure}[t]{0.33\textwidth}
 \includegraphics[width=\linewidth]{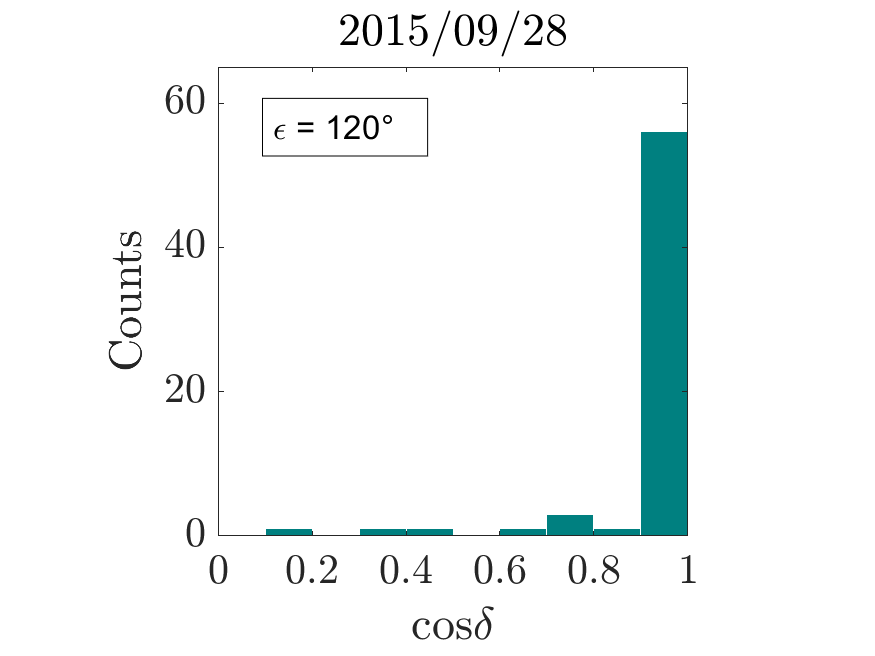}
  \end{subfigure}
  
  \begin{subfigure}[t]{0.33\textwidth}
 \centering
 \includegraphics[width=\linewidth]{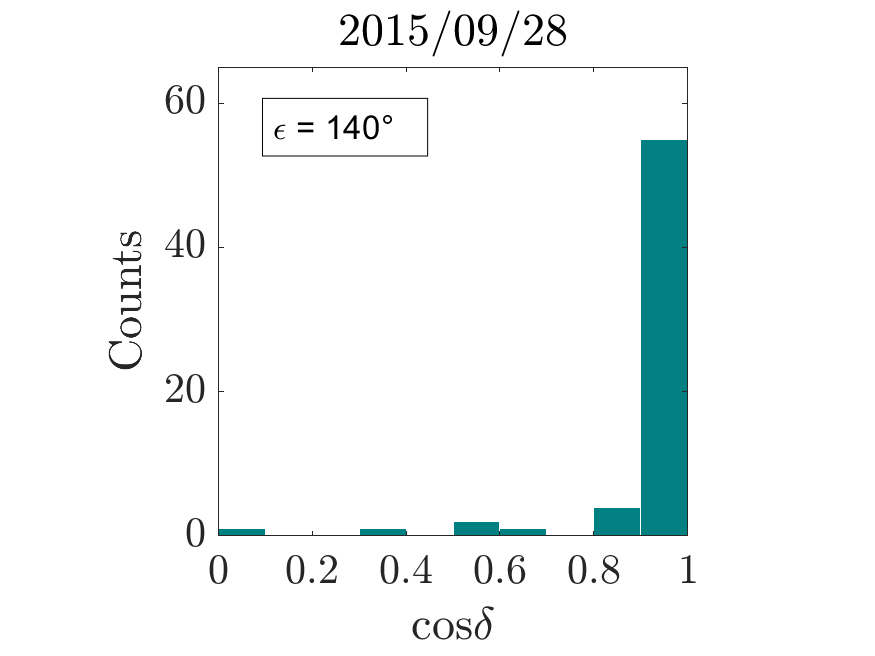}
  \end{subfigure}
 \begin{subfigure}[t]{0.33\textwidth}
 \includegraphics[width=\linewidth]{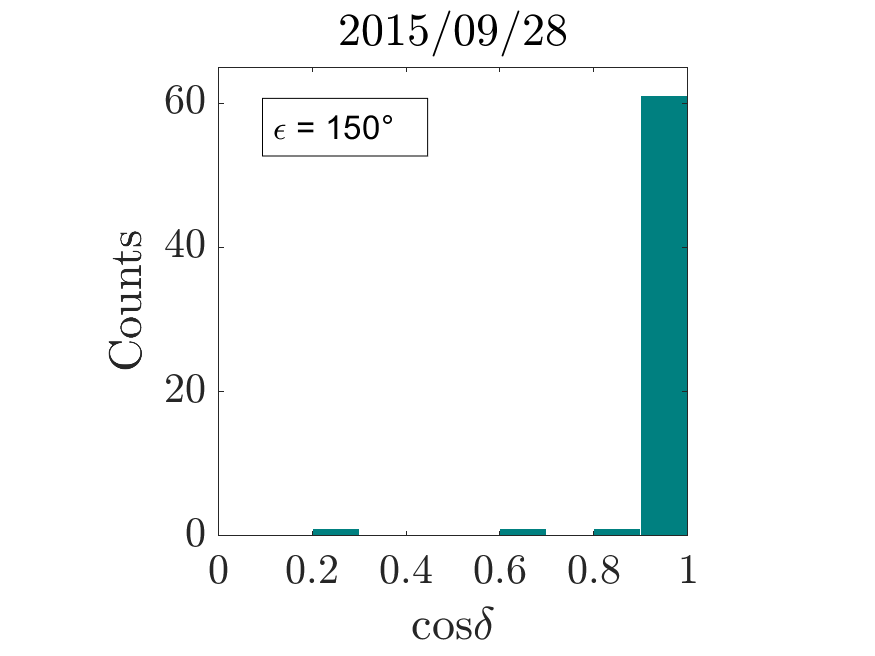}
  \end{subfigure}
  \begin{subfigure}[t]{0.33\textwidth}
 \includegraphics[width=\linewidth]{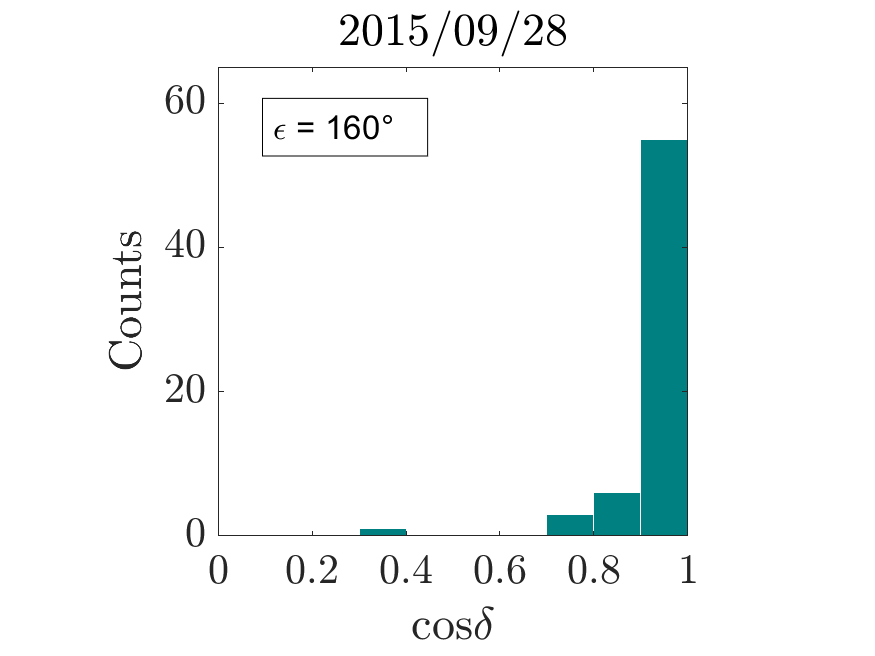}
  \end{subfigure}
     \caption{The histograms show the value of the cosine of the angle  $\delta$ between  the velocity vector of the particle on the image plane,  and the projected position vector of the nucleus  on the plane of the image, in the NAC reference system, for the set STP 075 GRAIN\_COLOR\_002.}
     \label{fig:radial75}
\end{figure*}

  \begin{figure}
    \centering
    \includegraphics[width = \linewidth]{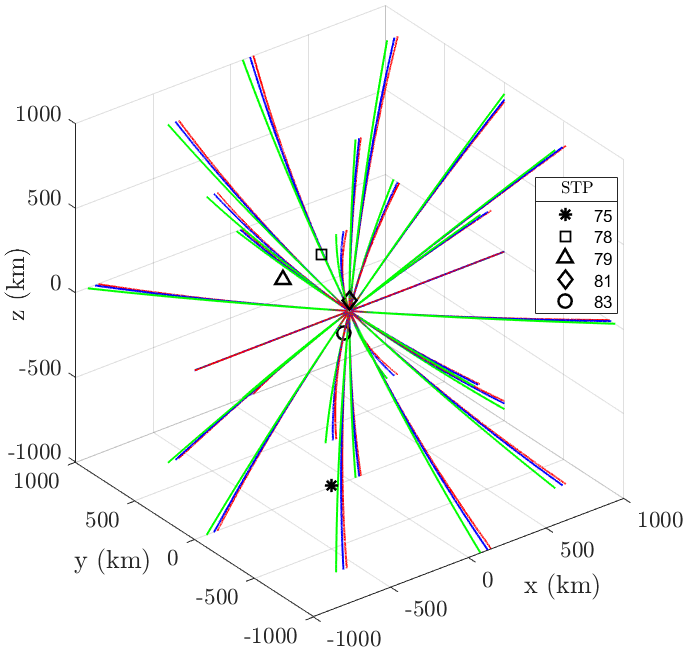}
    \caption{The figure shows the 3D trajectories of the dust particles ejected from the nucleus in all direction isotropically. The colour code is the following:  green for particles of 0.1 mm and v = 20 m/s, blue for particles of 1 mm and v = 5 m/s and red for particles of 10 mm and v = 1.5 m/s. Also the s/c position at each STP is shown.}
    \label{fig:traiettorie3D}
\end{figure}

 \subsection{Method}
To measure the direction of motion of the dust particles we evaluate the angles shown in Figure \ref{fig:technique}.
The angle $\delta$ is defined between  the velocity vector of the particle on the plane of the image, (defined by the track itself) and the projected position vector of the nucleus  on the plane of the image, in the NAC reference system,  while the angle $\beta$ is defined between   the projection of the velocity vector of the particle on the  plane of the image, and the projected position vector of the Sun.\\
The positions of the Sun and of the nucleus are derived using the SPICE Kernels \citep{acton1996} and eventually, the cosine of the angles  $\delta$ and $\beta$ are computed.
It is important to notice that the observational configuration changes within a single set of images. The elongation varies so the region of the coma investigated changes too. 

\subsection{Measurements}
We measure the direction of motion of the particles observed in the images taken in the post perihelion period ranging from September  to November 2015.
The results for   the set STP 075   are shown in Figure \ref{fig:radial75}. The results for the other set of images are reported   in   Appendix \ref{AppendixA}. Each histogram represents the values of the cosine of the angle $\delta$ that each particle detected in a single image forms with the nucleus position vector. The elongation ($\epsilon$) changes within a set, following the values illustrated in Table \ref{table:radiality}.\\
When the value of $\cos(\delta)$ is close to 1, the particle track, representing the projection of the motion on the plane of the image, is almost parallel to the direction vector of the nucleus. 
Our analysis shows that the majority of the particles have a value of $\cos(\delta)$ very close to 1, proving that the dust motion is mainly radial to the nucleus.
Therefore, the bulk of dust aggregates comes directly from the nucleus while the small deviations are probably generated by a population of dust inbound orbits around the nucleus \citep{ivano_BSE}.

 \subsection{Model}
We developed a simple model of dust dynamics in the region where dust is completely decoupled from the gas.
It is assumed that at the entrance in this region the dust particles have some initial velocity and further their motion is governed by the solar gravitational force, the solar radiation pressure and the nucleus gravitational force.  \\
We consider a static model in which the heliocentric distance of the nucleus and of the spacecraft are constant, justified by the short interval of time (hours) with respect to the heliocentric motion of the comet in which the ejection events occurred. 
The distance between the spacecraft and the nucleus is also constant.\\
The gravitational force of the Sun is defined as follows:
\begin{equation}
F_{G\odot} = \frac{G M_{\odot}}{r_{\odot}^2}m_p   
\end{equation}
where $m_p = \frac{4\pi \rho_d r^3}{3}$ is the mass of the particle, $\rho_d$ is the density of the particle and $r$ the radius. $r_{\odot} = (1.496\cdot10^{11}r_h$) is the heliocentric distance in metres.\\
The radiation pressure is:
\begin{equation}
 F_{rad} =  \frac{F_{\odot}}{ r_h^2} \frac{Q_{pr}\pi r^2}{c}
\end{equation}
where $F_{\odot} $ is the Sun radiation intensity at 1 AU  (corresponding to $F_{\odot} = 1360 $ W/m$^2$), $c$ is the speed of light, and Q$_{pr}$ is the scattering efficiency for radiation pressure (Q$_{pr}$ = 1 for total reflecting particles).\\
The gravitational force of the nucleus is:
\begin{equation}
F_{g} = \frac{G M_N}{r_c^2}m_p
\end{equation}
where   $M_{N} = 10^{13}$kg is the mass of the nucleus \citep{Paetzold2016}, and  $r_c$ is the distance of the particle from the nucleus.\\ 
We simulate an initial momentum spread isotropically from 10 km from  the surface of the nucleus, which is the distance at which the drag force   becomes negligible. The velocity is defined in the nucleus reference system as:
\[
\systeme*{v^{\prime}_x=v_0\sin{\theta}\cos{\phi}, v^{\prime}_y=v_0\sin{\theta}\sin{\phi}, v^{\prime}_z=v_0\cos{\theta}}
\]
where $v_0$ is the initial velocity value,    $\theta \in [0, \pi] $ is the polar angle  and $\phi \in [0, 2\pi]$ is the azimuthal angle.\\
We compute the trajectories travelled by dust particles of radius ranging from 0.1 mm to 10 mm. 
The model depends on the density of the dust $\rho_d$, the distance from the Sun $r_h$ and the initial velocity vector $v_0$.\\
We set $\rho_d = 800$ $kg/m^3$ \citep{fulle2016apj} and $r_h = 1.37$ AU as in the case of the set of images STP 075 taken  just after the perihelion.\\
The initial velocity is v = 20, 5 and 1.5 m/s, respectively, for particles of 0.1, 1 and 10 mm,  based on GIADA and OSIRS measurements of \cite{DellaCorte2016} and \cite{ott2017} and on the work of  \cite{stavro2017model}. We note that the dust velocity may depend on the direction from where the particle comes. Therefore, in our simple model with isotropic ejection of particles, we use some averaged velocities. 
The dust particles undergo the solar radiation pressure in different ways, depending on their size and mass. 
The smaller   the particle   the stronger   the effect of the solar radiation pressure.
At the same time, the small particles have higher initial velocities (due to acceleration by the gas).
The combination of these parameters result in an almost radial motion of all   particle size ranges considered. 
 Figure \ref{fig:traiettorie3D} shows the 3D trajectories of the particles of 0.1 mm, in green, of 1 mm in  blue and of 10 mm in red.
The Sun lies on the negative branch of the x-axis and the nucleus of 67P  is located at the origin of the axis. 
The black symbols represent the position of the s/c in the five epochs. 
The plane in which the camera is pointing is always perpendicular to the plane formed by the Sun, the nucleus and the spacecraft.
Following this model we find that the motion is almost radial in the inner part of the coma of the comet 67P for this range of particle sizes and initial velocities.
 
 \begin{figure} 
    \centering
            \includegraphics[width=\linewidth]{ 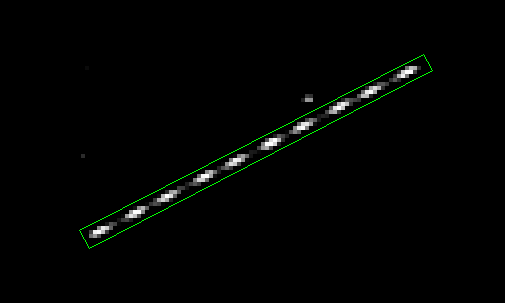}
        \quad
    \includegraphics[width=\linewidth]{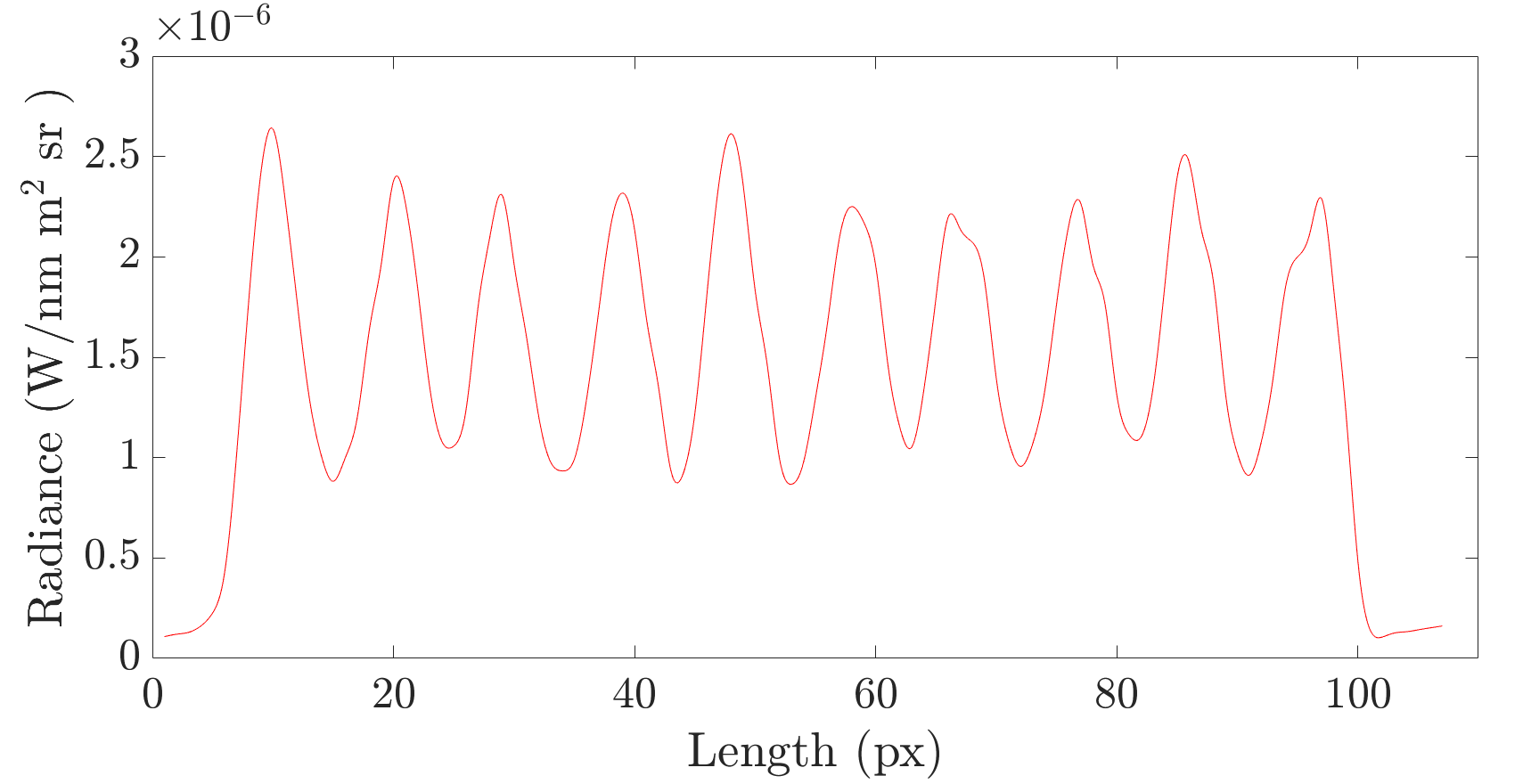}
  \caption {A rotating particle extracted from a single NAC image taken on 17 December 2015, STP 087, at UT 01:51:40.319 (top panel) and its  brightness profile  (bottom panel). The size of the image is 120x70 px.}
    \label{fig:light_curve1}
\end{figure} 

 \begin{figure} 
    \centering
    \includegraphics[width=\linewidth]{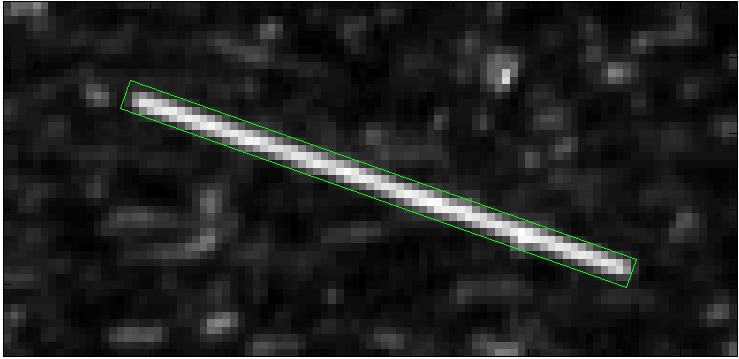}
           \quad
 \includegraphics[width=\linewidth]{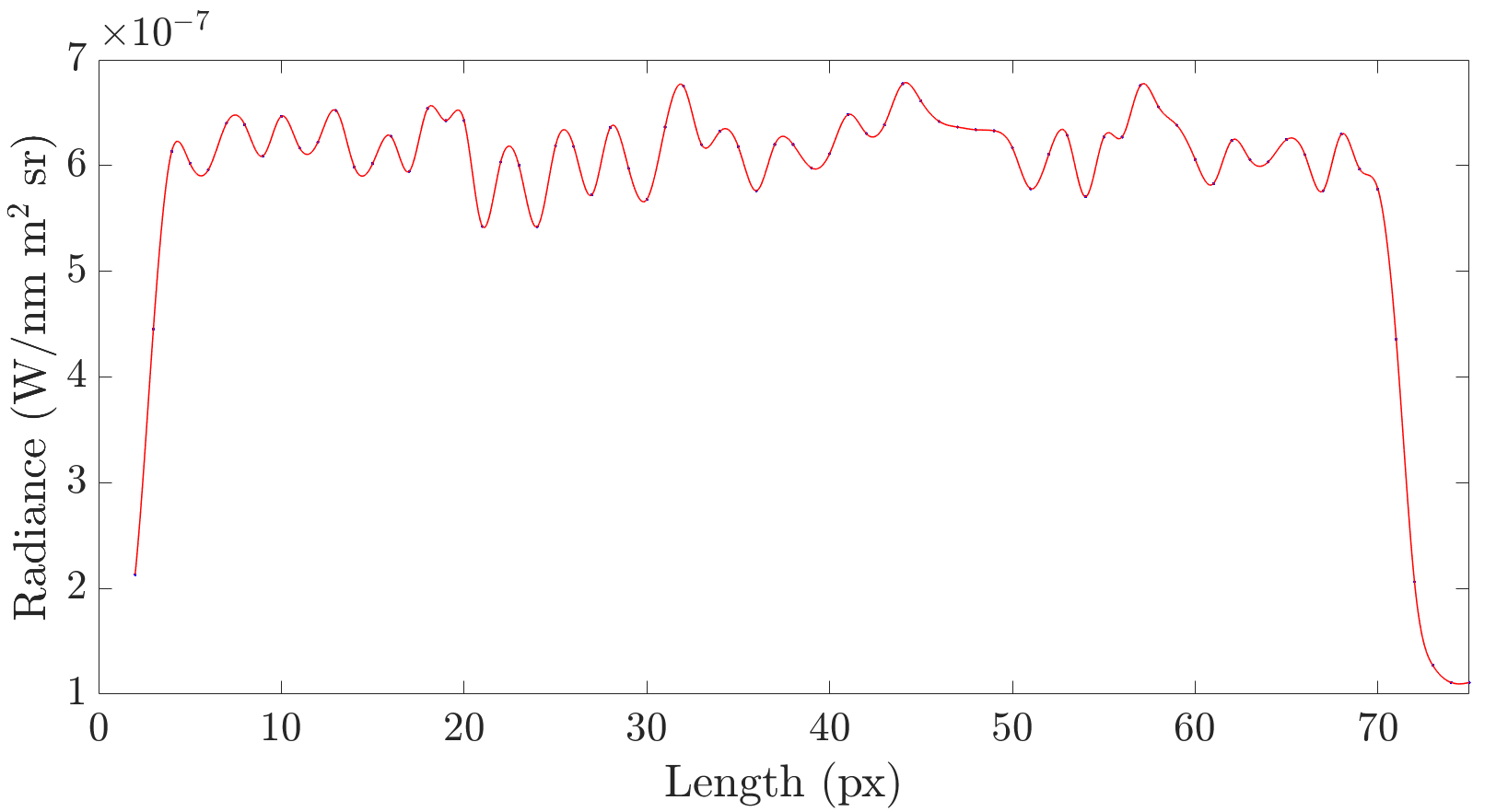}
  \caption {A particle that does not show evidence of rotation extracted from a single NAC image taken on 17 December 2015, STP 087, at UT 02:31:40.068 (top panel) and its  brightness profile  (bottom panel). The size of the image is 100x50 px.}
    \label{fig:light_curve2}
\end{figure} 

\section{Rotational state of dust particles} \label{rotational}
In this section, we present the analysis performed on   particles that show evidence of rotation.
We refer to the data set of Table \ref{table:periods}.
First, we measure  the rotational frequency of the particles, then we evaluate their flattening, giving some information about their shape. 

\subsection{Method}
We observe that some tracks show periodic variations in brightness, interpreted as rotational motion of the single dust particle. Figure \ref{fig:light_curve1} shows an example of light curve of a rotating particle. As a comparison, Figure \ref{fig:light_curve2} shows a light profile with  irregular radiance variations below the level of the background that can not be considered as evidence of rotation.
We evaluate each track to distinguish between  rotating particles and those that do not show evidence of rotation. 
We assume ellipsoidal particles with uniaxial rotation.
We consider the three rotational modes around the main axis both for the prolate and oblate particles.  
Depending on the modes of rotation the track generated by the passage of the particle in the field of view shows brightness variation or not.
For instance, an oblate spheroid rotating around its major axis perpendicular to the line of sight generates two peaks of brightness at each rotation, while if it rotates around its minor axis or around it major axis but parallel to the line of sight, it generates a homogeneous track. 
Consequently, the latter can not provide information about the particles shape.  
We use a criterion to classify the particles as rotating or not: the amplitude between the maximum and the minimum signal must be larger than the background signal. 
We discard those tracks that show only two peaks of brightness, in order to avoid  possible contamination from two different tracks \citep{FulleIvanovski2015}.
Overall, we  detect 1916 particles of which 344 show a periodical light curve that corresponds to 18$\%$ of the whole population.

 \begin{figure}
     \includegraphics[width= \linewidth]{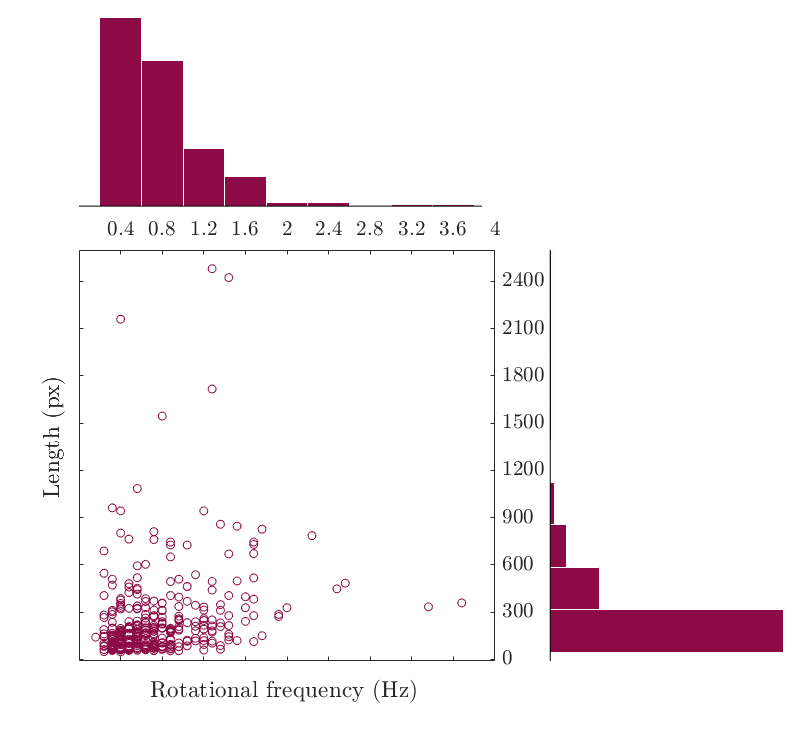}
   \caption{The figure shows the distribution of the particles in the parameter space  defined by the tracks  length (in px) and the rotational frequencies with their respective histograms.}
  \label{fig:length_peaks}
\end{figure}
 \begin{figure*}
\centering
\begin{tabular}{cc}
\includegraphics[width = 0.4\textwidth]{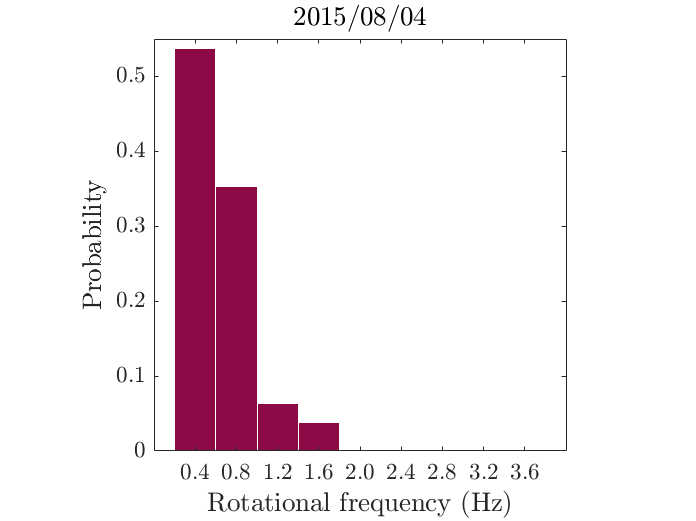}&
\includegraphics[width = 0.4\textwidth]{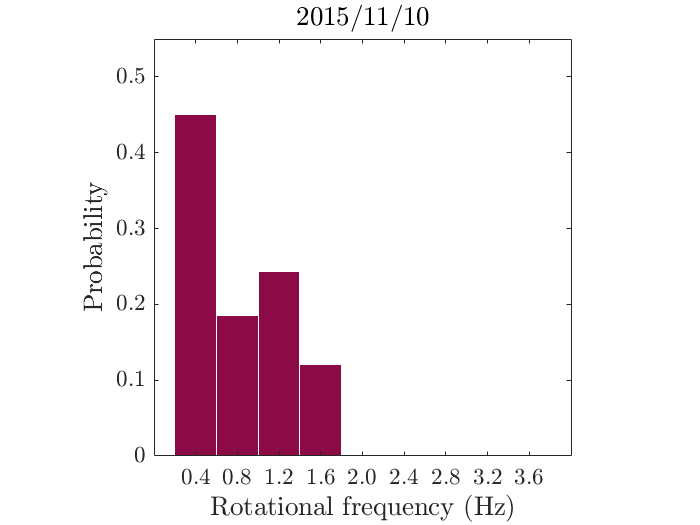}\\
\includegraphics[width = 0.4\textwidth]{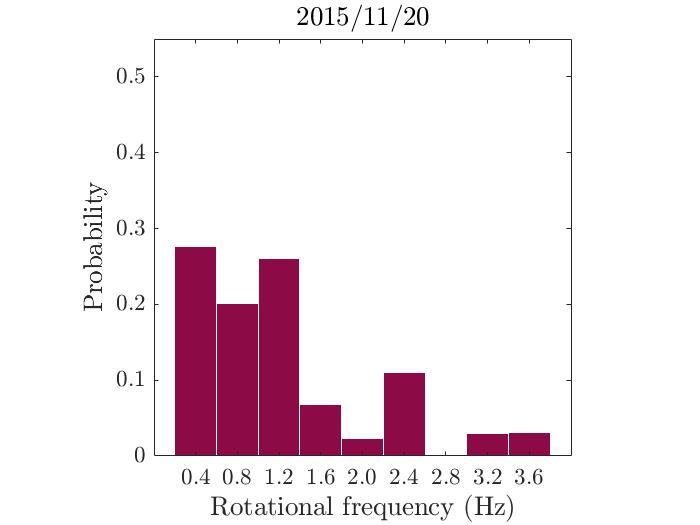}&
\includegraphics[width = 0.4\textwidth]{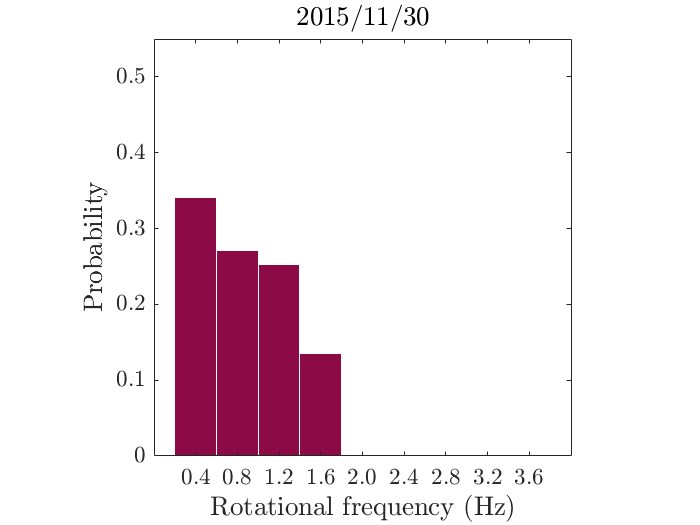}\\
\includegraphics[width = 0.4\textwidth]{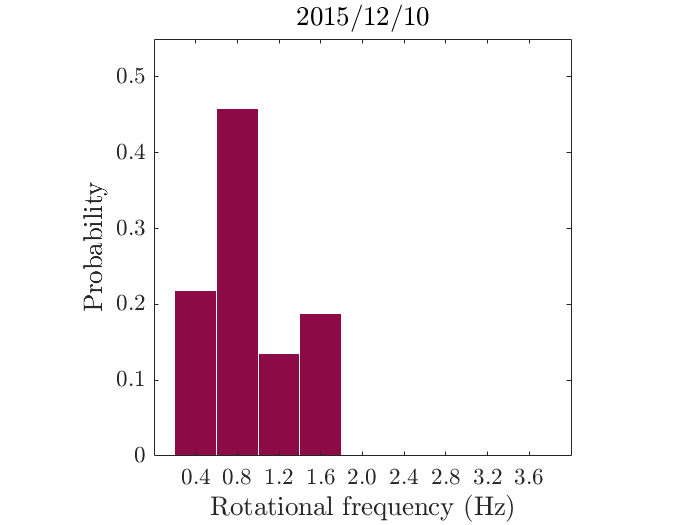}&
\includegraphics[width = 0.4\textwidth]{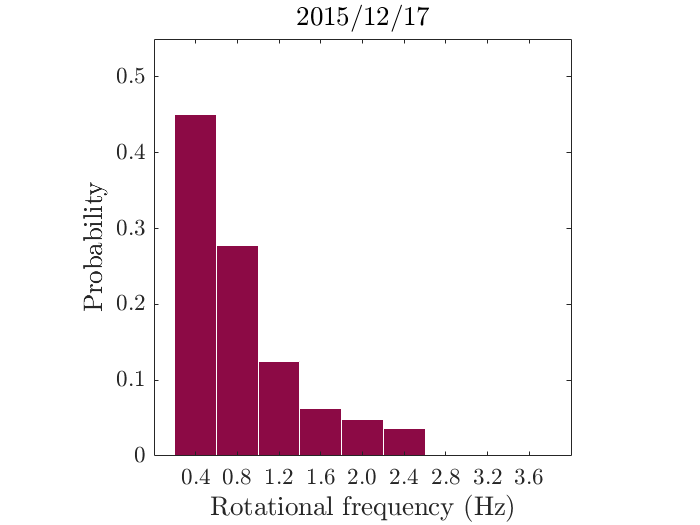}\\
\includegraphics[width = 0.4\textwidth]{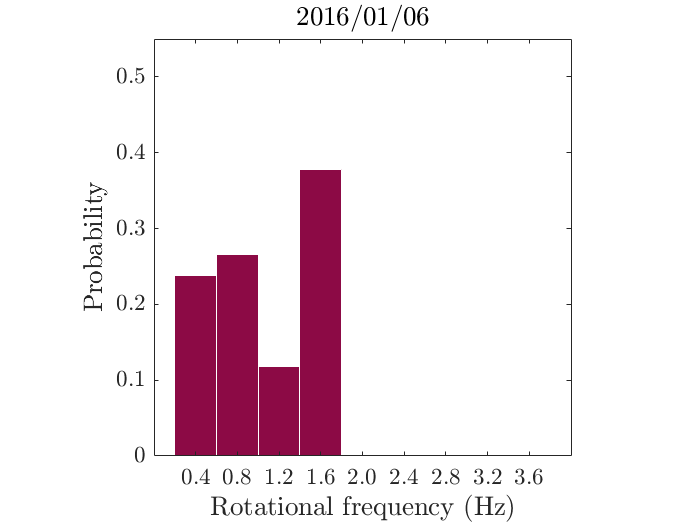}&
\end{tabular}
  \caption{The histograms represent the rotational frequency distribution of the particles detected in the images listed in Table \ref{table:periods}.}
  \label{fig:rotational_periods}
 \end{figure*}
 
 \begin{figure}
     \centering
     \includegraphics[width=\linewidth]{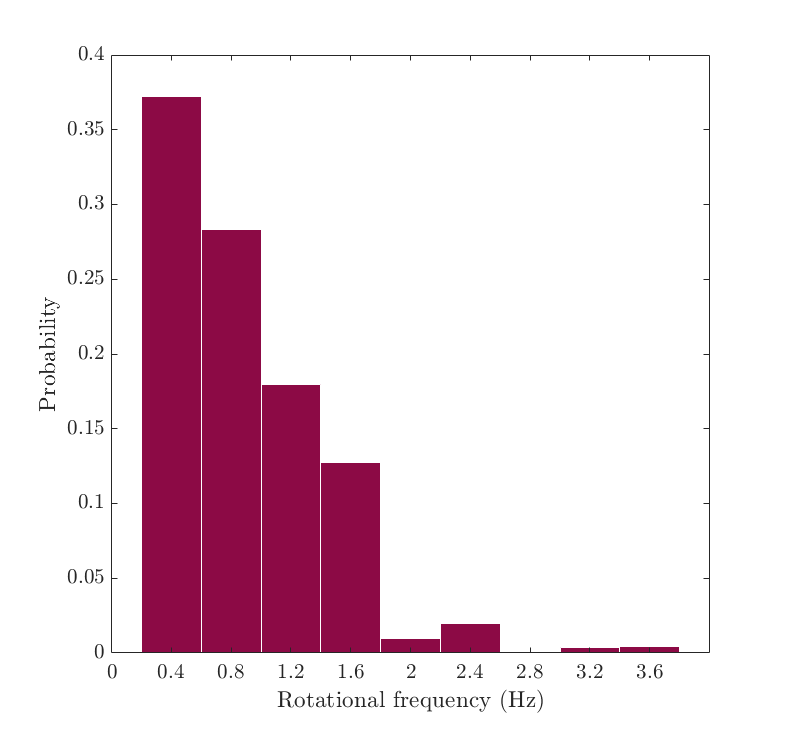}
     \caption{ The histogram  represents  the   distribution of the   rotational frequency of the whole sample of particles reported in Table \ref{table:periods}.}
     \label{fig:hist_periods}
 \end{figure}
 
\subsection{Rotational frequency}
 In approximation of uniaxial rotation of ellipsoidal particles with smooth homogeneous surface, the rotational frequency is straightforwardly computed dividing  the number of maximum brightness peaks in the light curves for the exposure time. 
It does not depend on the track length, neither on the track distance nor  the radiance, but only on the number of peaks.
Hereafter the rotational frequency refers to the brightness frequency.
Nevertheless, the statistics are affected by a bias due to the fact that only the longest tracks can provide fast rotations.
Therefore, in order not to underestimate the high frequencies, we
weight each track by its length, following \cite{FulleIvanovski2015}. 
The probability of obtaining a specific rotational frequency is given by the formula $p_n = f_n \cdot l_n / \sum_{i = 1}^{N}l_i$, where $f_n$ is the rotational frequency of the particle, $l_n$ is the length of the track and $N$ is the total number of tracks in the image. In Figure \ref{fig:length_peaks} the length-frequency parameter space is shown.
Furthermore, we correct for the projection effect.
 Since the trajectories are almost radial, as shown in the previous section, we can approximate the angle they form with the line of sight to the angle $\gamma$ shown in Table \ref{table:periods}. Therefore, the final length of a track (l$_n$) is equal to the length observed divided by the sine of the angle gamma.  
In Figure  \ref{fig:rotational_periods}  we report the distribution of the rotational frequencies weighted for the length of the tracks for the entire sample of particles detected in every single daily set.\\
All the images sets show a higher probability to find slow rotating particles rather than   fast ones.
Moreover, the decreasing trend is common to all sets of images  taken at different epochs and at different nucleocentric distances. This suggests that we are looking at a population of dust particles that did not undergo significant morphological and dimensional changes after their ejection from the nucleus.  
Therefore, a cumulative study of the data is allowed.\\
In Figure \ref{fig:hist_periods}, the distribution of the rotational frequencies of the whole set of images is merged in one single histogram.
 It results in a clear decreasing trend from $0.4 \pm 0.2$ up to a maximum of $3.6 \pm 0.2$ Hz, where the uncertainty corresponds to the bin sensitivity.
About   $40\%$ of the particles have frequencies between the minimum value of 0.24 Hz (corresponding to three peaks of brightness) and 0.56 Hz  (corresponding to seven peaks of brightness). \\
These values are in agreement with the ones obtained by \cite{FulleIvanovski2015} who found a comparable decreasing trend of the rotational frequency during the pre-perihelion period. 
This trend lets us suppose that the majority of the particles lies on the very left part of the histogram, which can not be examined since the  limited exposure time of our sample can only provide rotational frequencies above the minimum value of 0.24 Hz. Slower frequencies can be investigated looking at images with longer exposure time. Therefore, we expect that the particles do not rotate very fast in the inner coma, excluding a massive effect of fragmentation driven by their rotational motion. 
Overall, the tracks that show brightness variation are a percentage that range between 11-46\% of the average number of tracks observed in a single image (see Table \ref{table:periods}). 
Therefore, all  tracks without brightness variation could be representative of either very slow rotators, with periods larger than the exposure time of the image, or particles that are rotating exposing always the same face. 

\subsection{Flattening}\label{flattening}

\begin{figure}
    \centering
    \includegraphics[width=\linewidth]{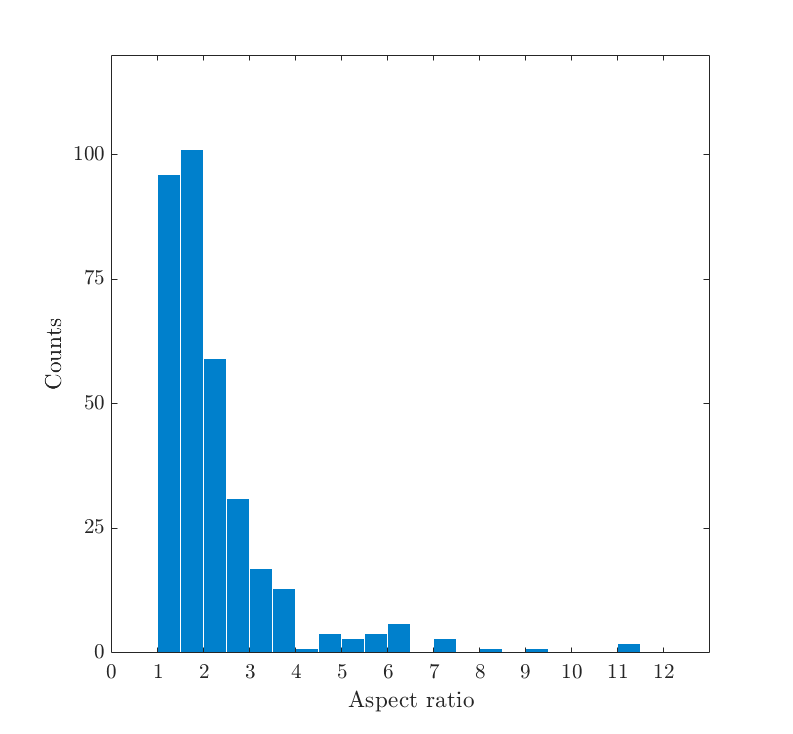}
    \caption{The histogram shows the ellipsoidal particles aspect ratio distribution   listed in Table \ref{table:periods}.}
    \label{fig:axis_ratio}
\end{figure}

\begin{figure}
    \centering
    \includegraphics[width=\linewidth]{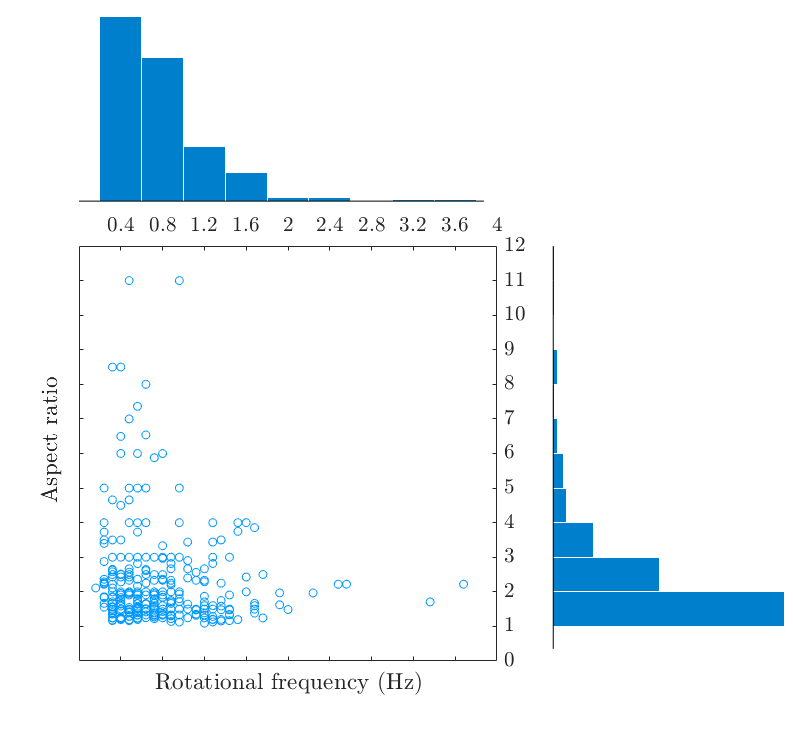}
    \caption{The figure shows the distribution of the particles in the parameter  space  defined by the particle aspect ratios and the rotational frequencies with their respective histograms.}
    \label{fig:axis_ratio2}
\end{figure}

The morphology of the dust of the order of hundreds of microns in 67P's coma was revealed by the images of the Micro-Imaging Dust Analysis System (MIDAS \citep{Riedler2007})  and COSIMA  \citep{mannel2016,langevin2016}.
The dust particles show irregular shapes,  they can be either fluffy aggregates or compact particles \citep{carsten2019}. 
To give an estimate of their flattening, we 
evaluate the aspect ratio of particles via measurements of the  maximum to minimum brightness ratio of the light curves, 
that represents the projected area of the aggregates on the plane of the image assuming uniaxial rotation and homogeneous composition.
We decide to approximate these shapes with an ellipsoid of revolution, i.e. spheroid, which represents, as a first approximation, an irregular aggregate   with two main cross sections on the plane of the image.   
Spheroids are also used as dust shape models by \cite{FulleIvanovski2015}, \cite{stavro2017model, stavro2017giada}  \cite{moreno2018} to represent the dynamics of the dust in the coma.
The estimated aspect ratios are always larger than 1 (which is the
value for a   sphere), because we cannot discriminate between  oblate and prolate spheroids. 
It means that a value equal to 2, for instance, could be attributed to a prolate spheroid but also to an oblate spheroid with aspect ratio 0.5.
In Figure \ref{fig:axis_ratio}  the distribution of the aspect ratio values of the whole sample of particles is shown. 
The majority of the population lies between 1.1, for almost spherical particles, and 2.
Nevertheless, the aspect ratio reaches values up to 11, revealing that the particles 
might be also very elongated \citep{Fulle2017}.\\
Following the results of \cite{FulleIvanovski2015} we might suppose a higher percentage of oblate spheroids rather than prolate. 
\cite{stavro2017giada} also found that oblate-like particles are the best candidates to reproduce GIADA data before perihelion. \cite{olga2020} show that porous particles of an oblate-like shape oriented with its largest area facing the laser beam are necessary to reproduce the minimum of the OSIRIS phase functions at a phase angle of about 100 degrees, as observed by \cite{ivano_phase}.\\
In Figure \ref{fig:axis_ratio2}, we show the rotational frequency as a function of aspect ratio.  
The majority of the particles fill homogeneously an area defined by values of aspect ratio between 1 and 2 and with frequency between 0.24 and 0.56 Hz.
Therefore, for those particles, it is not possible to infer a relation between rotational frequency and shape. Furthermore, we recognise two populations that deviate from the majority: one with high rotational  frequency and low values of aspect ratio so more spherical particle, the other   with high values of aspect ratio, up to about 11, which corresponds to very elongated particles with relatively slow rotational frequency. Indeed more elongated particles are expected to rotate slower than the spherical ones, and this is linked to their  inertia momenta \citep{stavro2017giada}.
So, if the majority of the population is composed by slow rotators we might expect more aggregates with high elongation.
 
\section{Photometry} \label{photometry}

In this section, we measure the photometry of the  particles in the images  taken in the period August 2015 - January 2016 (see Table \ref{table:periods}). Then, we simulate the expected spectral radiances of the dust particles in the cometary coma with a scattering model.
We compute the radiance of the single particles, changing a wide range of parameters such as the composition, size, heliocentric distance, and spacecraft distance. Finally, we compare them with the particles spectral radiances obtained directly from the images to estimate their size.

 \subsection{Scattering simulations}
 
\begin{figure}
    \centering
    \includegraphics[width=\linewidth]{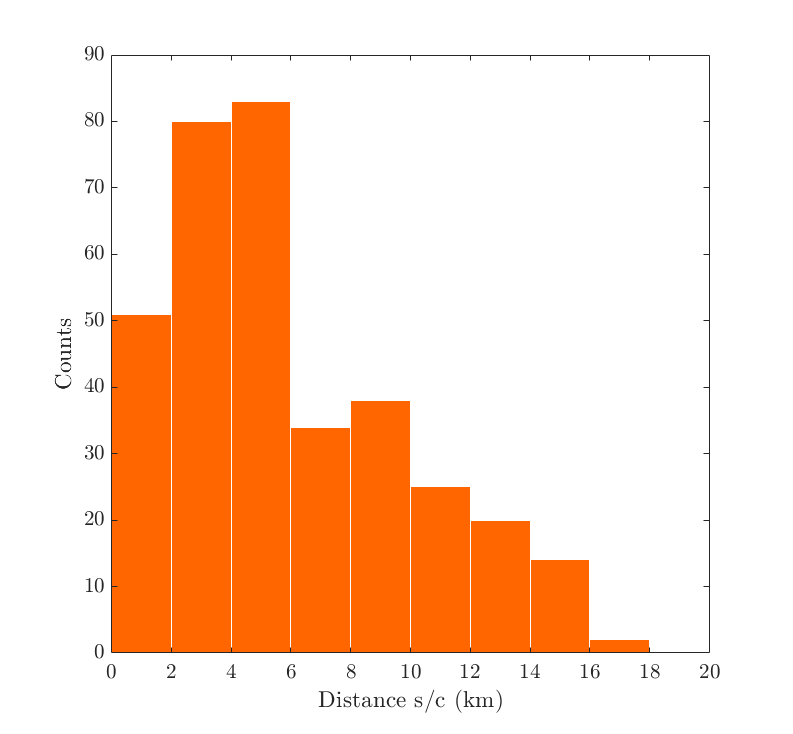}
    \caption{In the figure  the    distribution of the  particle  estimated distances from the spacecraft is shown. }
    \label{fig:distance_sc}
\end{figure}

We use a scattering code in order to synthetically reproduce the photometry of the  dust particles.\\
It takes into account a wide range of parameters such as the distance from the spacecraft and the Sun, the size of the particles, their phase angle, density and the type of material they are made of, evaluated  through the refractive index.\\
 The code computes the amount of light scattered by each dust  particle towards the CCD by using the Mie theory for  spherical particles \citep{MieCode}. The software  is optimise to consider arbitrary   large size  parameters and to give a first estimate of the particles size.
 Input parameters of the code are the phase angle of the dust, the density, and the refractive index of the material. 
 We fix the phase angle at 90 degrees and the density at 800 kg/m$^3$ \citep{fulle2016apj}. 
 We select two characteristic compositions of the dust particles represented by the silicates, with refractive index m$ = 1.6 + i10^{-5}$ \citep{munoz_allende,frattin_granada}, and by the organics, with refractive index m$ = 1.3 + i10^{-2}$
 \citep{frattin_granada}.\\ 
  This choice is based on the observations of the 67P dust provided by the instruments on board the Rosetta s/c.
 COSIMA mass spectrometer has confirmed the presence of abundant high-molecular weight organic matter, nearly 50 \% in mass \citep{bardyn2017,fray2017}. Half of the mass of each dust particle consists of carbonaceous material with a mainly macromolecular organic structure; the other half being mostly composed of non-hydrated silicate minerals. 
 GIADA has found that the measured densities of the 67P pebbles are consistent with a mixture of (15 $\pm$ 6) \% of ices, (5 $\pm$ 2) \% of Fe-sulphides, (28 $\pm$ 5) \% of silicates, and (52 $\pm$ 12) \% of hydrocarbons, in average volume abundances \citep{fulle2016apj,Fulle2017}.
 \cite{frattin} have found hints on the presence of three major groups of dust particles, based on the analysis of their spectral slope between [535–882] nm obtained from OSIRIS images. They found that the steepest spectra   may be related with organic material, the spectra with an intermediate slope, can be a mixture of silicates and organics and the flat spectra  may be associated with a high abundance of water ice \citep{frattin}.   
Following these results, organics and silicates seem to be good representatives of the dust particles composition in the coma of 67P.\\
In Table \ref{tab:scattering_model} we report the expected spectral radiances for different combinations of the code parameters.
We simulate particles with a size of 0.1, 1, 10 mm, at distance from the spacecraft that ranges from 1 km to 20 km, following the results shown in Figure \ref{fig:distance_sc}. 
We fix the heliocentric distance at  1.3, 1.7 and 2.0 AU to best represent the observations.
The resulting spectral radiances have values between $10^{-3}$ and $10^{-10}$ W/(m$^2$ nm sr). 
They are strongly dependent on the heliocentric distance, showing the highest values at perihelion and decreasing as the comet moves away from the Sun, and are brighter for silicate particles than for the organic ones.

 \subsection{Results}
 
\begin{figure*}
\centering
\begin{tabular}{cc}
\includegraphics[width = 0.4\textwidth]{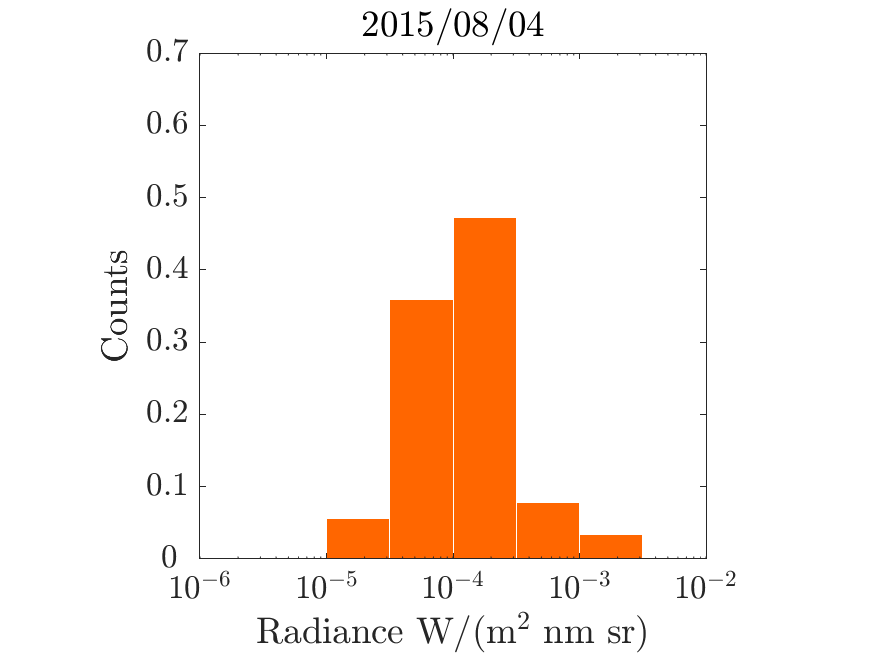}&
\includegraphics[width = 0.4\textwidth]{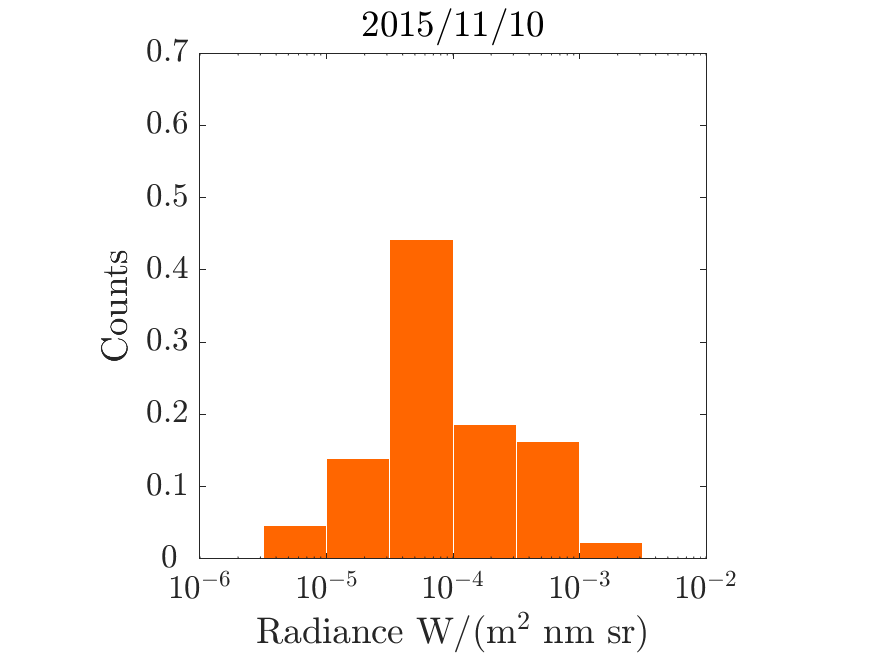}\\
\includegraphics[width = 0.4\textwidth]{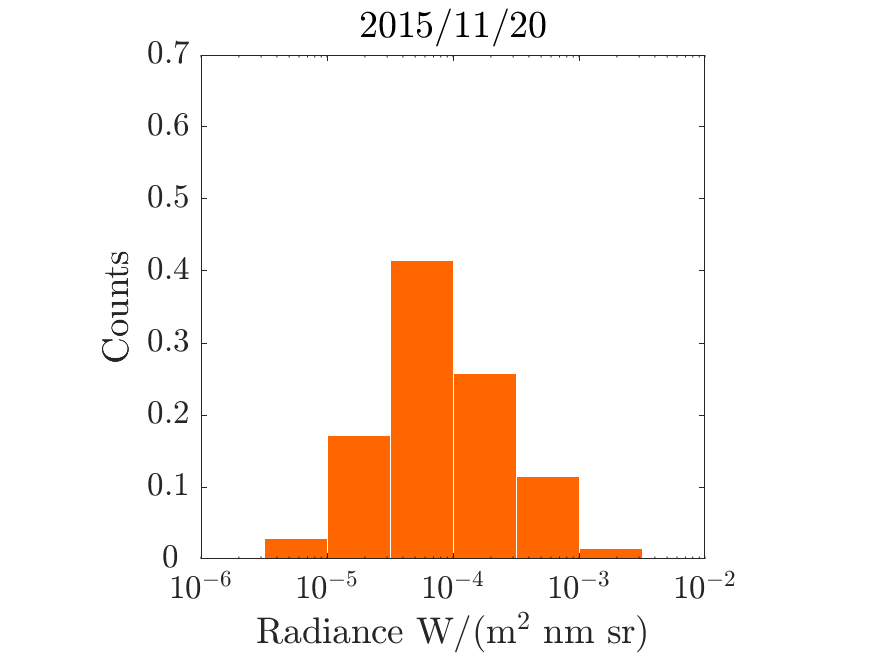}&
\includegraphics[width = 0.4\textwidth]{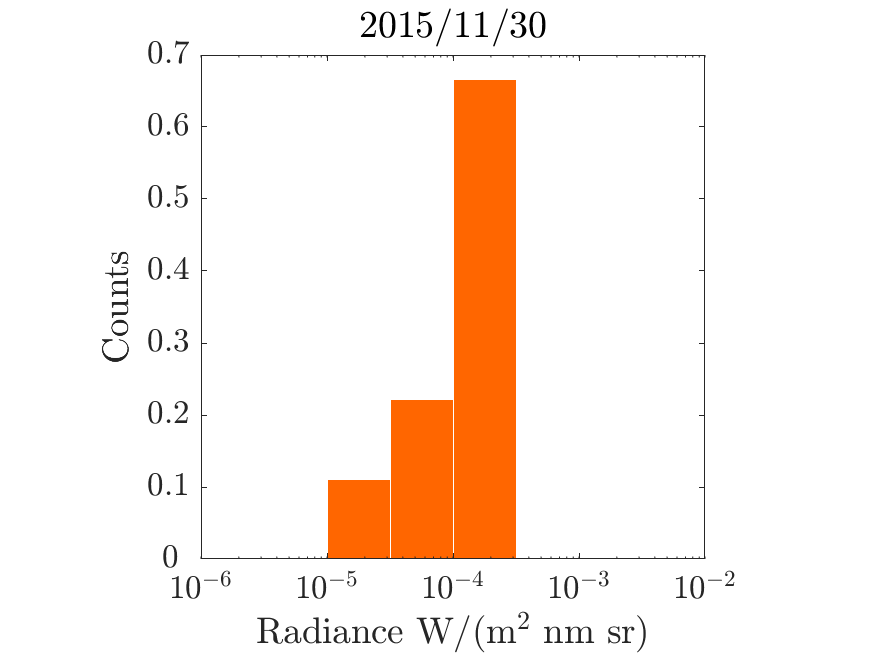}\\
\includegraphics[width = 0.4\textwidth]{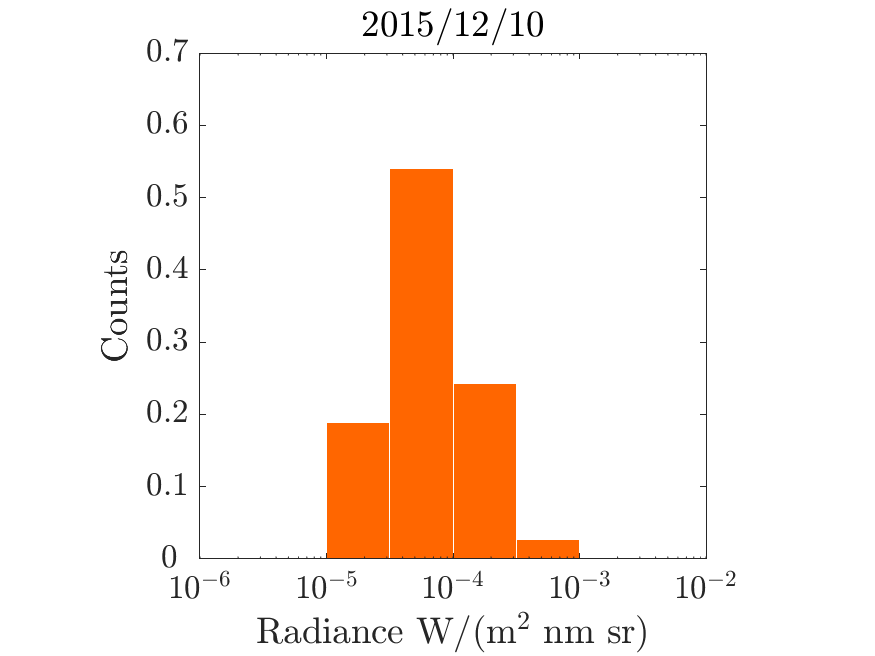}&
\includegraphics[width = 0.4\textwidth]{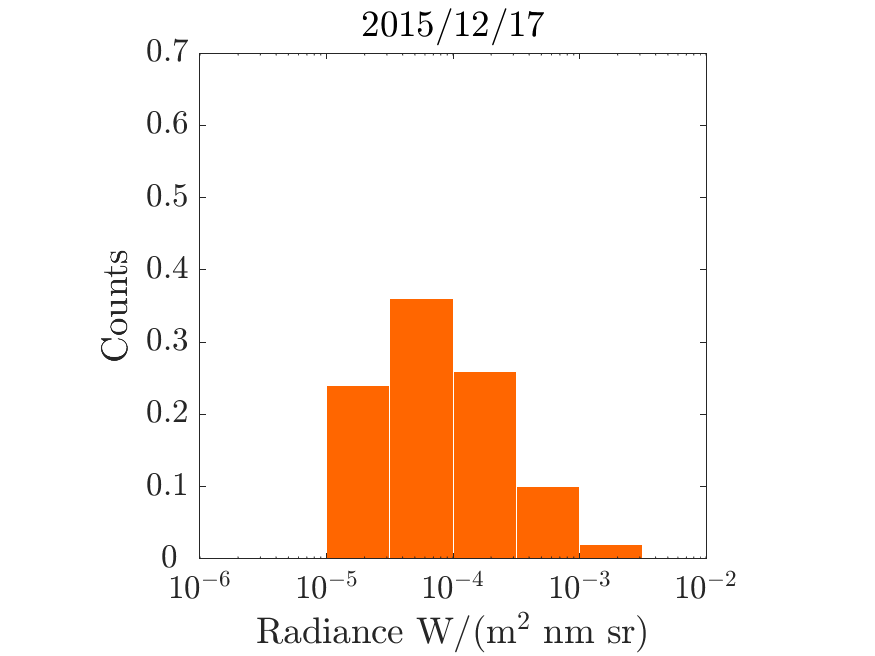}\\
\includegraphics[width = 0.4\textwidth]{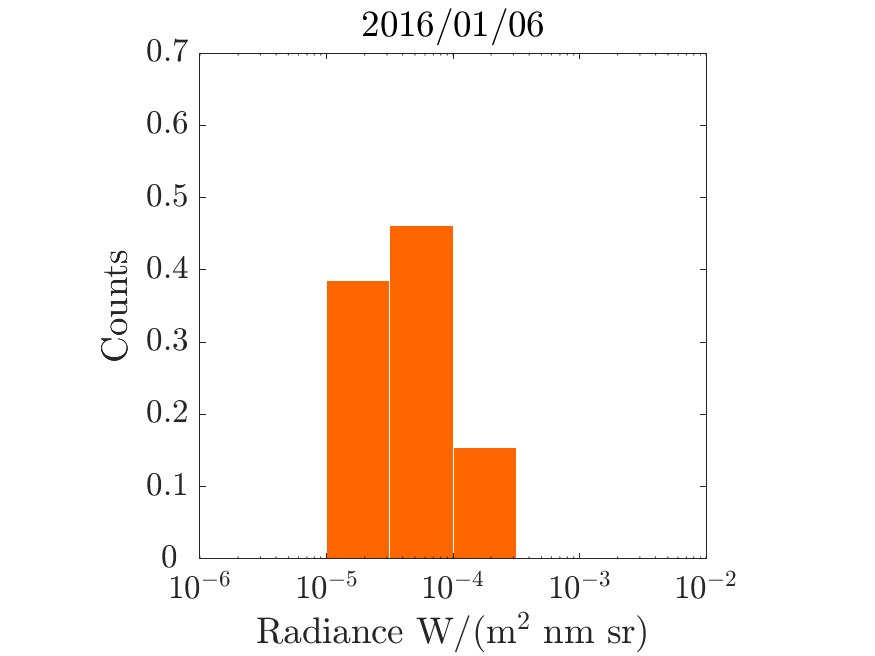}&
\end{tabular}
  \caption{The graphics show the histograms of the dust particle spectral radiances for each epoch from the data set illustrated in    Table \ref{table:periods}. The sample  consists in the same rotating particles analysed in Section \ref{rotational}. 
The counts are normalised to 1. The spectral radiance is corrected for the phase function and it is in units of  W/(m$^2$ nm sr).}
  \label{fig:fluxes}
 \end{figure*}

 We perform the photometry of the rotating dust particles analysed in the previous section and reported in the dataset of Table \ref{table:periods}. 
 We evaluate the radiances of the whole track computing the sum of the signal of each pixel divided by the exposure time.
In Figure \ref{fig:fluxes} we see the distribution of the particle spectral radiances for each epoch. The values range between $10^{-6}$ and $10^{-3}$ W/(m$^2$ nm sr), but the large majority is between   $ 10^{-5}$ and $10^{-4}$ W/(m$^2$ nm sr).  The   uncertainty associated to the radiometric calibration  is less than 1$\%$ \citep{ceci_filtri}.\\
In order to give an estimate of the effective size (based on brightness) of the  aggregates, we compared them with the  simulated values. 
For the calibration, we use the data of \cite{ott2017}.
We choose 2 aggregates at about 3-4 km from the spacecraft, at 1.25 AU from the Sun, with a cm-size  as a reference (precisely number 8 and 9 of their Table 3). 
We find that the values of the spectral radiance, corrected for the phase function \citep{ivano_phase},  are respectively of  $7.02 \pm 0.02 \cdot 10^{-4}$ and $5.15 \pm 0.02 \cdot 10^{-4}$ W/(m$^2$ nm sr), where the uncertainty is the standard deviation of the background.
These results are strongly compatible with the simulations, whether we   consider either organics or silicates.
As a result, we  find that the most populated values of spectral radiance correspond to mm and cm-size.
The faintest value of   radiance might be compatible with a particle size of the order of 0.1 mm  (if very close to the s/c, i.e. 1 km) in   both  the cases of silicates and organics. 
Simulated particles smaller than mm have spectral radiances fainter than  $10^{-7}$  W/(m$^2$ nm sr), a value that is outside the range obtained in the observations.
Therefore, our analysis of rotational particles suggests that the majority of the dust aggregates we observe might likely have sizes that range from mm to cm.

 \subsection{Particles Distance and Size}
\begin{figure}
    \centering
    \includegraphics[width = \linewidth]{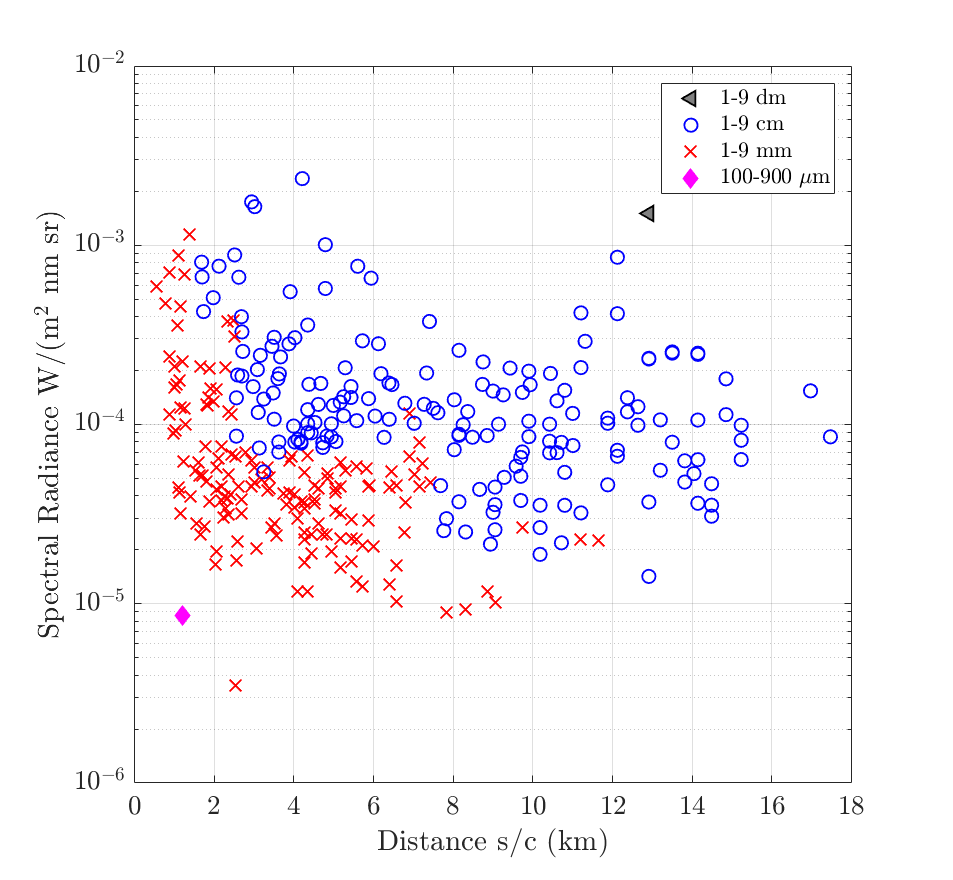}
    \caption{The plot illustrates the distribution of the dust particles in the parameter space defined by their spectral radiances and estimated distances from the spacecraft. The black triangle is a dm-sized particle, the blue circles are the cm-sized particles, the red crosses are the mm-sized particles while the magenta diamond is a $\mu$m-sized particle.}
    \label{fig:diagram}
\end{figure}

\begin{figure}
    \centering
    \includegraphics[width = \linewidth]{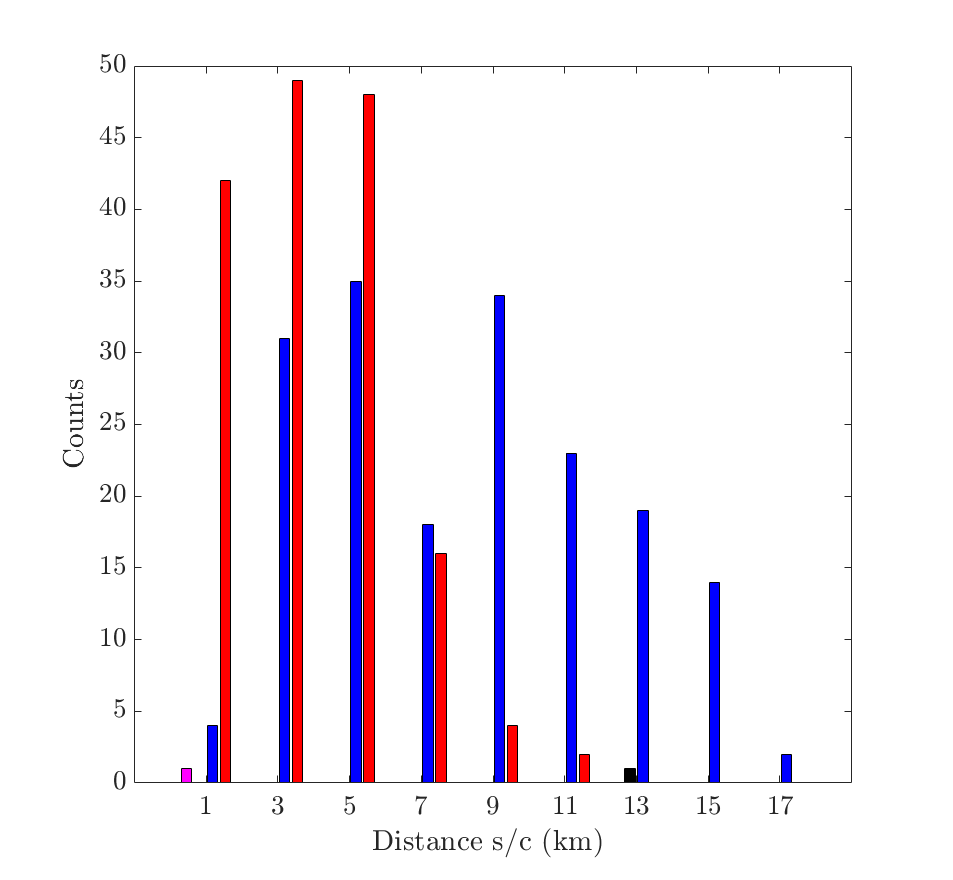}
    \caption{The histogram represents the distribution of the dust particles of different size along the line of sight. The black bars represent the dm-sized particles, the blue ones the cm-sized particles while the red ones the mm-sized particles.}
    \label{fig:size_distribution}
\end{figure}

We estimate the minimum size of the observed rotating particles, based on the radiance computed with the scattering model.
The faintest measured spectral radiance has value equal to  3$\cdot$10$^{-6}$  W/(m$^2$ nm sr). 
We use the results of the scattering model to determine the minimum possible size corresponding to this radiance. 
To do this we set the heliocentric distance and the spacecraft distance of the particles at their minimum values, respectively $r_h$ = 1.3 AU and $d_{s/c}$ = 1 km. In fact, the smaller is the distance of the particles from the Sun and from the spacecraft and the higher is their radiance. In this way, the possibility to detect a small particle is maximised. We see that the radiance equal to 3$\cdot$10$^{-6}$ matches to a particle smaller than 1 mm (to which corresponds a radiance equal to 3$\cdot$10$^{-5}$) but larger than 0.1 mm (to which corresponds a radiance equal to 3$\cdot$10$^{-7}$). Therefore, the particles that we observe in this work are at least of the order of 0.5 mm. This is a lower limit computed in case of particular conditions therefore we expect the majority of the particles to be larger than that limit.
The same size range was studied by \cite{ott2017} who found particles from $1.5 \cdot 10^{-3}$ m up to 0.32 m with an average value of 0.05 m.
They were able to measure also the speed of those particles, finding that the 90\% of them had values between 0 - 5 m/s with statistic mode equal to 1.25 m/s (see their Figure 9). 
The detection of rotating particles is affected by a bias that facilitates the observation of larger and slower particles. 
Indeed, very small and faint particles should approach the spacecraft very close to be detected and consequently they would move out of focus.
Moreover, particles of the order of 0.1 mm or even  smaller,  as those measured by GIADA \citep{rotundi2015, DellaCorte2015} might reach high velocity and the tracks they generate extend along more pixels with a consequent decrease of intensity that complicates the detection.
Therefore, based on the scattering model and on the previous assumptions, we have reasons to think that the tracks observed in these set of images are generated by particles likely larger than 1 mm.
Furthermore, the track signal must be high to be able to reveal a brightness variation, thus excluding the fainter particles from the sample.
Following this assumption, we use the velocity found by \cite{ott2017} for particles in the same range  of size.
We approximate the average velocity at which the observed particles move inside the coma with their statistic mode, $v = 1.25$ m/s.
Moreover, we suppose that within the same image the longest tracks represent on average closer particles, while the shorter ones are particles far away from the spacecraft. \\
The length and the number of rotations depend also on the mass and the size of the particles. We expect that the smallest and lowest density particles reach higher velocity and angular velocity than the large and dense ones \citep{stavro2017model}. In the latter paper, the authors show that this could change if   large particles of low density and small particles of high density are studied. 
In Section \ref{section:stavro} we show the results of dust modeling of particles with the same  derived dynamical characteristics obtained by this analysis of the observed light curves.
Nevertheless, a first investigation can be done, to provide hints about those rotating particles.\\ 
We use the length of the tracks to measure the distance of the single particles.  
In the approximation that the particle velocity distribution mode is a good indicator of the average speed of the observed particles,
 it is straightforward that the average length covered by a dust particle in 12.5 seconds of exposure time is of $v\cdot t= 15.625$ m.
 Therefore, it is possible to construct the following spatial scale: $v \cdot t /l = 15.625/l $  m/px , where $l$ is the length of the tracks in pixel.
 On the other side, the spatial scale of the NAC camera is 1.86  m/px  at 100 km from the spacecraft. 
Equalling the two scales leads to the following expression:
\begin{equation}
 1.86 : 100 = \frac{v\cdot t}{l}  : d_{s/c}
\end{equation}
\noindent from which  d$_{s/c}$, the distance of the particles from  the spacecraft is immediately computed.
The results are shown in Figure \ref{fig:distance_sc}, where the histogram represents the average distance distribution.
Eventually, the particles analysed in this work, which are observed with the Orange filter at $\lambda$ = 649 nm, with an exposure time of 12.5 seconds seem to reach distance from the spacecraft of less than 20 km. \\
 Ultimately, we compute the size  of the rotating particles.  
We compare the observed values of the spectral radiance  of each dust particle with the values obtained with the scattering model at proper heliocentric and spacecraft distances, reported in Appendix \ref{AppendixB}. Each combination of these three parameters (r$_h$, $d_{s/c}$ and the radiance) corresponds to a specific size. 
In Figure \ref{fig:diagram} we show the parameter space composed by the observed spectral radiances and distances of the dust population listed in Table \ref{table:periods}. 
The distribution of the particles along the line of sight size is shown in Figure \ref{fig:size_distribution}. 
The cm-sized group seems to be the most populated.
These results are consistent with the direct measurements  of \cite{ott2017}, that found that the bulk of particles has sizes of centimetres.

%
\begin{table*}
	\centering
\begin{tabular}{lc p{9cm}}
\hline
   {\bf{Input parameter}} 	& {\bf{Parameter value}} 	& 	{\bf{Comments}} \\  
\hline   
 \multicolumn{3}{ c }{{\bf{Particle parameters}}} \\    
\hline
Particle size [m]  & $1.0\times 10^{-3}$  $-$ $1.0\times 10^{-1}$ & Observational constraint derived from this study \\
Particle density [kg/m$^3$] & $800$ & Constrained by the GIADA data \citep{2016MNRAS.462S.132F,Fulle2017}\\
Spheroid aspect ratio, ${a/b}$ & $2/3$ and $3/2$ & Observational constrain derived as the most relevant in this study \\
Initial particle orientation [deg] & $22.5 - 67.5{^\circ}$ & Compatible with the one used to reproduce particle speeds measured by GIADA \citep{stavro2017giada} \\
Particle temperature, $ T_d $ [K]:   & 300  & \cite{fulle2020a}\\
\hline 
   \multicolumn{3}{ c }  {{\bf{67P/CG parameters}} } \\   
\hline
Radius, $R_N$ [m]   & $2.0\times 10^3$ & \citet{Sierks2015} \rule{0pt}{11pt} \\
Mass, $m_N$ [kg]   & $1.0\times 10^{13}$ & \citet{Paetzold2016}\\
 \multicolumn{1}{l} {Total gas production [mol/s]}&$10^{28}$&{\citet{2016MNRAS.462S.491H}}\\
\hline  
\end{tabular}
\caption{{The dust model initial parameters used in this work.}}
	\label{tab:tab_param}
\end{table*}

\section{Dust dynamical simulations of rotating non-spherical particles} \label{section:stavro}
\par 
The 3D+t non-spherical cometary dust model \citep{FulleIvanovski2015, stavro2017model, stavro2017giada} has been the first one applied to Rosetta data and allowed to obtain those dust physical parameters that were not possible to measure by the on-board instruments due to the operational limitations or instrumental design. 
First, the model was used in \cite{FulleIvanovski2015} to interpret the brightness variation of the tracks in OSIRIS images taken in October 2014. 
In the present work, we apply the model to datasets of NAC images from July 2015 to January 2016 considering the physical conditions present in 67P coma during this period. 
We compute the particle rotational frequencies and terminal velocities. We provide a plausible scenario for interpreting the data assuming initial parameters that can lead to the rotational frequencies derived from the observations. 

\subsection{Model and setup}
We consider homogeneous spheroids with aspect ratio 1.5 - the most probable obtained by the observations in this study.  
The two main forces acting on the dust particles in closer vicinity to the nucleus are the aerodynamic force and gravity. The analysis of the images show mainly radial outflow (Section \ref{radial}). 
We consider that the gas approximation for calculating the aerodynamic force is the Euler one for an expanding ideal gas \citep{landau_book}, which is less time and computationally consuming. The calibration of the physical parameters of these gas solutions is obtained using the GIADA data and the gas production rates reported in \cite{2016MNRAS.462S.491H}.  
We shortly review the equations governing the motion of given dust particle:       
   \begin{equation}
     \begin{array}{l}
      m_d \frac{d^2 \vec r}{dt^2} = m_d \frac{d \vec V_d}{dt} = \vec F_N + \vec F_a \\ 
      \\
       \vec F_N =  - G\frac{ m_N m_d}{r^3} \vec r \\ 
       \\
       \vec F_a =  - \int \left( p \vec n + \tau \left[ (\vec V_r \times \vec n) \times \vec n \right]/ |\vec V_r\times \vec n| \right) ds,    \\                                          
    \end{array}  
    \label{eqn_base}
   \end{equation}
where $\vec{r}$ and $\vec V_d$ are the radius-vector and the velocity of the center of mass of the particle in  the cometocentric non rotating frame $x$, $y$, $z$; $m_d$ is the particle mass; $m_N$ is the nucleus mass,  $\vec V_r$ is the gas velocity relative to the elementary surface element of the particle, and  $\vec F_N$ and $\vec F_a$ are  
the gravity and the aerodynamic forces; $p$, $\tau$ are the gas pressure and the shear stress on the elementary surface element of the particle with area $ds$ (for the exact free molecular expressions, see \cite{Shen});              
$\vec n$ is the outward unit vector normal to the elementary surface element, and $G$ is the gravitational constant. 
Since the minimal collisional mean free path of the gas molecules on the surface is of the order of tens of metres, i.e. much bigger than the considered particle sizes, the flow over the particles may be considered as free molecular  \citep{stavro2017model}.\\
We integrate equations (\ref{eqn_base}) and the Euler dynamical and kinematic equations \citep[for details]{Landau1969} for large distances \citep{stavro2017giada}. The gravitational, $\vec M_N$, and aerodynamic, $\vec M_a$, torques necessary for computing the Euler dynamical equations are given by:      
   \begin{equation}
   \begin{array}{l}
       \vec M_N = - \int G m_N \tilde \rho  \frac{\vec l_{s} \times \vec r}{r^3}d\Gamma \\         
       \\                                                                   
         \vec M_a = - \int \vec l_{s} \times \left( p \vec n + \tau \left[ (\vec V_r \times \vec n) \times \vec n \right]/|\vec V_r\times \vec n| \right) ds \\
    \end{array}  
    \label{eqn_torque}
   \end{equation}
 where $d\Gamma$  is the elementary volume of the particle, $\vec l_{s}$ is the radius-vector towards $ds$ or $d\Gamma$ from the center of mass of the particle, and $\tilde \rho$ is the dust specific mass.
 \par
For the goals of this study, we use the dust parameters constrained by the observations in this paper and the physical parameters of 67P as its mass, gas production and surface temperature in the considered period. Table \ref{tab:tab_param} summarises the simulation parameters in two different sets. The first set includes the physical properties of the simulated particles. The second set reviews the 67P measured physical parameters used for the gravity force computation. In particular, we assume a particle bulk density of 800 kg m$^{-3}$ \citep{Fulle2017} and non zero initial velocity (see the next subsection). We assume that the particles do not contain any volatiles nor do they fragment, i.e. keep their shape and mass during the whole motion. 
%
   \begin{figure*}
   \centering
     \includegraphics[width=\columnwidth]{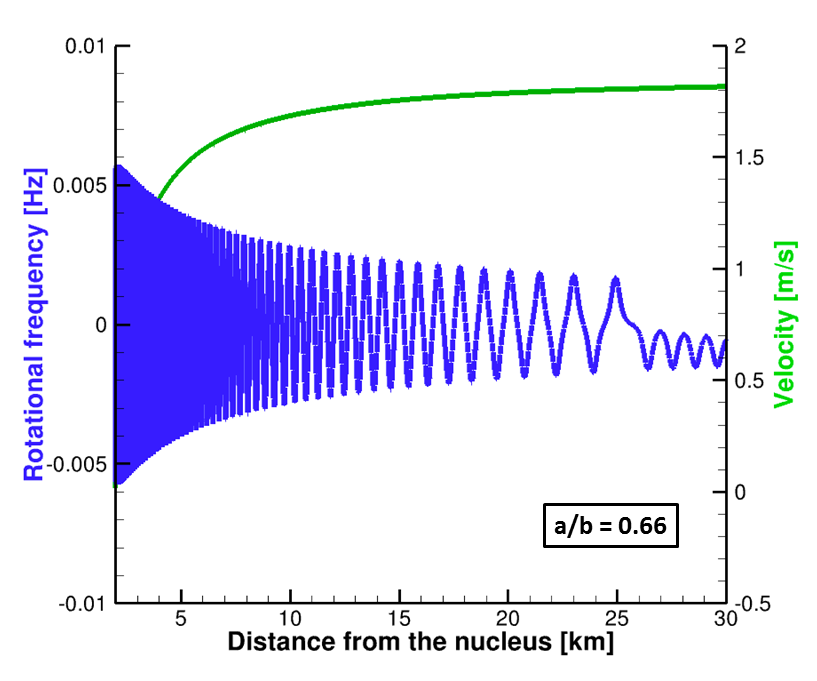}
     \includegraphics[width=\columnwidth]{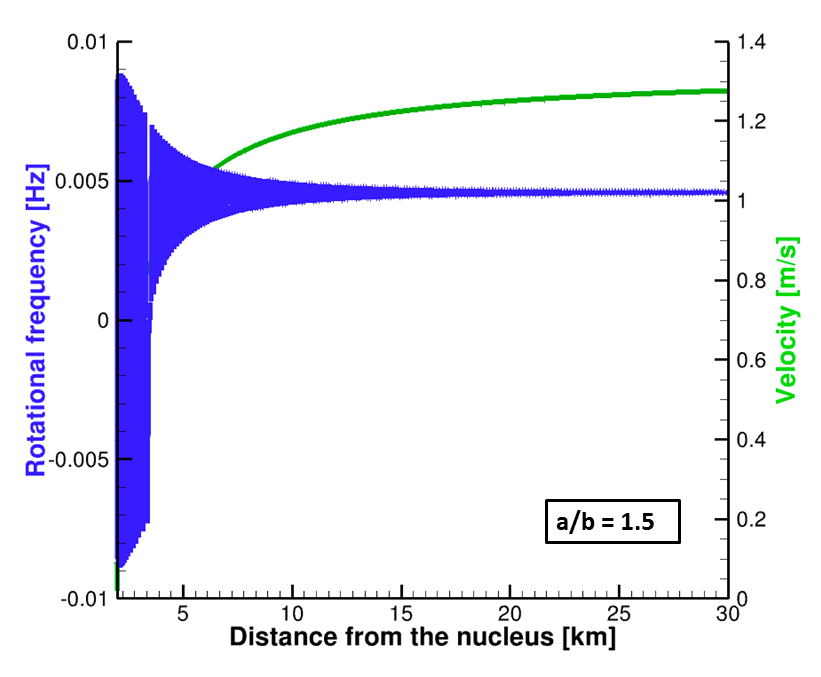}
     \caption{Particle velocity (green lines) and rotational frequencies(blue lines) in two cases: spheroid with aspect ratio $a/b= 0.66$ (left) and spheroid with aspect ratio $a/b= 1.5$ (right). Both spheroids are with size $b= 1$ cm in radius and bulk density 800 kg/m$^{-3}$ is considered. The complete list of the particle parameters are given in Table \ref{tab:tab_param}.}     
     \label{fig:torque}
   \end{figure*}     

 \subsection{Computed rotational frequencies and particle speeds} 
We compute the velocities of spheroidal particles with different aspect ratio $a/b = $2/3 and 3/2 (oblate and prolate) ranging from $b$ = 0.01 to 1 m. Figure \ref{fig:torque} shows the computed particle velocities  and rotational frequencies of spheroids of the same size ($b$ =1 cm) and density (800 kg m$^{-3}$) as found as the most probable candidates derived in the present study from the analysis of the OSIRIS data. These velocities and rotation frequencies are close to those in \cite{stavro2017giada} found for particles of the same size, moving in the same gas field, but with density 1000 kg m$^{-3}$ and $a/b$ = 2.0. In the present simulations the gas production rate is 100 times greater than used in \cite{FulleIvanovski2015} (due to the variation of  the gas production from pre-landing period in October 2014 to perihelion in August 2015). The results show that the considered particles remain slow rotators (< 0.01 Hz).
\par
Figure \ref{fig:hist_periods} shows that the majority of the observed particles have frequency $\sim$ 0.4 Hz. The most brightness frequency in October 2014 data was 0.3 Hz, while during and post perihelion is 0.4 Hz. 
 If we compute the most probably rotational frequency considering both the shorter exposure time 12.5 s ($\sim$ 2/3 of the one used in the measurements during 12-13 and 19-20 October 2014 (Figure 4, \cite{FulleIvanovski2015})  and the most probable frequency of those periods  i.e.\ $\sim$ 0.3, we obtain as a most probable frequency $\sim$ 0.4 i.e.\ the same as the one derived in our measurements analysis. The latter fact evidences that the most probable rotational frequency has not changed when approaching the Sun.
\par
Since the perihelion data analysis can not disentangle the prolate of oblate shapes, we perform simulations for both shapes , i.e. $a/b$ = 2/3 and 3/2. The oblate particle obtain higher rotational frequency than the prolate particle, but it is still lower than derived from the observations. Among the possible explanations could be uncertainty of the particle detachment process (i. e. presence of initial velocity and/or torque and/or rocket force due to particle sublimation). Therefore we perform simulations for these spheroids assuming that they have initial velocity which corresponds to the maximum possible ejection velocity of a particle with the same volume and mass estimated for 67P nucleus in \cite{fulle2020a} (assuming complete transfer of the gas thermal energy at the detachment into the initial kinetic energy of the particle). This initial velocity is of few centimetres per second and does not affect the resulting rotational motion in our simulations. Owing to the large uncertainty in the values to be adopted for the initial torque and possible outgasing from the particle we do not try to address this in the current work. A preliminary study shows that the high initial rotational frequencies (> 0.4 Hz) can remain unchanged during the particle motion through the coma.

\section{Summary and Conclusions}   \label{conclusion}
In this work, we have analysed a collection of images of the inner coma of the comet 67P/Churyumov-Gerasimenko with the aim of describing the physical and dynamical properties of the single dust particles to provide constraints for theoretical 3D coma models.\\
In Section \ref{radial} we performed a dynamical analysis of the particles in the inner coma of 67P. 
We measured the direction of motion  directly from the images revealing  highly radial trajectories with respect to the nucleus. This suggested that the material originated from the nucleus and that all deviations from the radiality might be interpreted as a small population of bound orbiting objects.\\
In Section \ref{rotational} we focused on a group of particles that showed evidence of rotational motion. 
The sample represented about the 18\% of the whole population, while the remaining showed homogeneously bright tracks. 
We measured a rotational frequency with a decreasing trend up to a maximum value  of $\nu = 3.6 \pm 0.2$ Hz. 
It was not possible to detect particles with  values of rotational frequency below 0.24 Hz since the exposure time of the images was not long enough to identify slow rotating objects. 
The decreasing trend of the frequency distribution and the fact that the sample represented only a small part of the overall dust particles population suggested that the majority of them rotated slower than the  group studied in this work. 
Consequently,  no fragmentation of the aggregates was expected. 
The light curves analysis provided  also the aspect ratio of the single dust particles in the approximation of ellipsoidal shape, contributing to characterise them. 
The majority of the population had values that range between 1 and 2, revealing a moderate level of flattening.
Nevertheless, the aspect ratio reached values quite high, up to 11, revealing the particles might be also very elongated.\\
In Section \ref{photometry} we performed the photometric analysis of the same samples of particles used in the previous section. 
We measured their fluxes and, by comparing them with  simulated ones we derived their  size.  
We took into account two types of composition, i.e. silicates and organic, and we computed the radiance produced by different sizes  at different distances from the camera with a Mie scattering model.
The direct comparison of the simulated fluxes with the observed ones   provided a size distribution of the particles that ranged from the value of millimetres up to the decimetres.
Following those calculations, it seemed that the camera was not able to detect particles smaller than millimetres  (except one case of a particle of the order of 0.1 mm) since they should have had a flux fainter than the background.\\
These observational measurements worked as constraints for the theoretical dynamical model used in Section \ref{section:stavro} to compute the velocity and rotational frequencies of the dust particles.
The results of the simulation showed that particles with the same characteristics of the observed ones remained slow rotators ($\nu$< 0.01 Hz). Therefore, the simulations confirmed that the majority of the particle population should have had a low rotational frequency. 
The explanation to characterise the motion of the fast rotating particles still misses. A possible hint could be related with the uncertainty of the particle detachment processes that might transmit them an initial rotational velocity.\\

\noindent {\bf Data Availability}
All the data underlying this paper are available at https://www.cosmos.esa.int/web/psa/rosetta.
 


\bibliographystyle{mnras}
\bibliography{Observational_constraints_to_dust_dynamics_paper_accepted.bib} 

\section*{Acknowledgements}

OSIRIS was built by a consortium of the Max-Planck-Institut fur Sonnensystemforschung in Gottingen, Germany; CISAS-University of Padova, Italy; the Laboratoire d'Astrophysique de Marseille, France; the Instituto de Astrofisica de Andalucia, CSIC, Granada, Spain; the Research and Scientic Support Department of the European Space Agency, Noordwijk, The Netherlands; the Instituto Nacional de Tecnica Aeroespacial, Madrid, Spain; the Universidad Politechnica de Madrid, Spain; the Department of Physics and Astronomy of Uppsala University, Sweden; and the Institut fur Datentechnik und Kommunikationsnetze der Technischen Universit at
Braunschweig, Germany. The support of the national funding agencies of Germany (DLR), Italy (ASI), France (CNES), Spain (MEC), Sweden (SNSB), and the ESA Technical Directorate is gratefully acknowledged.\\
GIADA was built by a consortium led by the Univ. Napoli
Parthenope  INAF- Oss. Astr. Capodimonte, in collaboration with
the Inst. de Astrofisica de Andalucia, Selex-ES, FI and SENER. GIADA was managed and operated at INAF-IAPS. GIADA was funded
and managed by the Agenzia Spaziale Italiana, IT, with the support
of the Spanish Ministry of Education and Science MEC, ES.\\
 We thank the ESA teams at ESAC, ESOC, and ESTEC for their work in support of the Rosetta mission.\\
This research has been supported by the Italian Space Agency (ASI) within the ASI-INAF agreements I/032/05/0, I/024/12/0 and ASI 2020-4-HH.0.\\
Fernando Moreno acknowledges financial support from the Spanish Plan Nacional de Astronomia y Astrofisica LEONIDAS project RTI2018-095330-B-100, project P18-RT-1854 from Junta de Andalucia, and the Centro de Excelencia Severo Ochoa Program under grant SEV-2017-0709.\\
Elisa Frattin  acknowledges funding from Italian Ministry of Education, University and Research (MIUR) through the "Dipartimenti di eccellenza” project Science of the Universe.


\begin{landscape}

  \begin{table} 
\centering
    \begin{tabular}{cccccccccc}
\hline 
MTP & STP  & Activity  &  Epoch & Start time& $\Delta t$(hh mm ss) & $r_c$(km)& $r_h$(AU) & $\alpha(^{\circ})$ &  $\epsilon(^{\circ})$   \\
\hline \hline
021& 075 & GRAIN\_COLOR\_002& 2015.09.28 & 14:42:32.150& 01.04.00 & 1220 & 1.37 &  50&   90-160 \\
\hline 
021& 078 &GRAIN\_COLOR\_004& 2015.10.18& 02:51:30.857&01.28.00& 437 & 1.47 &  65&    70-150 \\
\hline  
022& 079&GRAIN\_COLOR\_001&2015.10.23&20:04:28.191 &01.28.00 &389 & 1.51 &  63&   70-130   \\
\hline  
022& 081&GRAIN\_COLOR\_002& 2015.11.06&  20:05:59.163& 01.31.00& 244 & 1.61 &  63&   55-140   \\
\hline 
023& 083&GRAIN\_COLOR\_001& 2015.11.19& 13:11:24.629 &01.36.00 &126 & 1.70 &  74&  50-130   \\
 \hline \hline
     \end{tabular}
     \caption{ OSIRIS data set used in Section \ref{radial} to measure the particles direction of motion and their alignment with nucleus and Sun directions. MTP: medium time planning of the dataset. STP: short time planning of the dataset. Activity: the activity name of the sequence. Epoch: image acquisition day. Start time: image initial acquisition  time. $\Delta$t: range of time between the first and last image acquisition. r$_c$: nucleocentric distance. r$_h$: heliocentric distance. $\alpha$ : phase angle as shown in Figure \ref{fig:geo_81_GC2}   (i.e. angle between spacecraft and Sun as seen by the comet). $\epsilon$: elongation of the Sun (i.e. angle between the camera pointing direction and the Sun as seen by the s/c). The exposure time of each image is 12.5 s.} 
    \label{table:radiality}  
\end{table}

\begin{table}
\centering
    \begin{tabular}{cccccccccccccccccc}
\hline
MTP & STP & Activity  &     Epoch & Start time &$\Delta$t (hh mm ss)&  $r_c$(km) & $r_h$(AU) & $\alpha(^{\circ})$ &  $\epsilon(^{\circ})$ &$\gamma(^{\circ})$ & \#rot & \#im & \# tot & \%rot& $f$(Hz) & a/b & bkg    \\
\hline \hline
 019 & 067   & GRAIN\_COLOR\_002 & 2015.08.04 & 14:20:31.208 & 01.15.19& 238 & 1.25 & 90  &   135 & 45 & 89 & 8 &599 & 25 &0.4 &1.5& $3\cdot10^{-7}$ \\
\hline 
022 & 081   & GRAIN\_TRACK\_003 & 2015.11.10 & 13:32:44.361 & 03.30.01 & 202 & 1.61 & 62  & 137 & 19 & 43 & 12& 284 & 22& 0.4  &1.5& $2\cdot10^{-7}$ \\
\hline  
023 & 083    & GRAIN\_TRACK\_001 & 2015.11.20 & 18:29:26.800 & 03.20.01 &142 & 1.70 & 87 & 122 & 29 & 70 & 6  & 355 &30 &  0.4 &1.9& $2\cdot10^{-7}$   \\
\hline 
 023 & 084     & GRAIN\_TRACK\_002 &2015.11.30 &17:56:25.051 & 02.20.01 & 111 & 1.70 & 90 &  120  & 30 & 9 & 7 &  107 & 11& 0.4 &1.1& $1\cdot10^{-7}$ \\
\hline  
 023 & 086    & GRAIN\_TRACK\_003 &2015.12.10& 17:32:23.417 &03.50.00 &99 & 1.86 & 89  &   76 & -15 & 74 & 10& 301& 41 & 0.8  &2.4& $4\cdot10^{-7}$ \\
\hline 
024 & 087   &GRAIN\_TRACK\_001 &2015.12.17 &01:53:01.775 & 04.30.00 &100 & 1.9 & 90 &   106& 16 & 50& 12 & 198 & 46& 0.4   &1.3& $1\cdot10^{-7}$ \\
\hline
 024 & 090     &GRAIN\_TRACK\_003 &2016.01.06& 11:13:43.828 &02.40.00 &84 & 2.06 & 90 &   75  & -15 & 13 & 9  & 72 & 38& 1.6 & 1.6& $2\cdot10^{-7}$\\
\hline\hline
     \end{tabular}
     \caption{
     OSIRIS data set used in Section  \ref{rotational} for the analysis of dust rotational motion. MTP: medium time planning of data acquisition. STP: short time planning of data acquisition. 
     Activity: the activity name of the sequence. Epoch: image acquisition day. Start time: image initial acquisition  time.
      $\Delta$t: range of time between the first and last image acquisition. r$_c$: nucleocentric distance. r$_h$: heliocentric distance. $\alpha$: phase angle as shown in Figure \ref{fig:geo_81_GT3}. $\epsilon$: elongation of the Sun (i.e. angle between the camera pointing direction and the Sun as seen by the s/c). $\gamma$: angle between the comet and the pointing direction as seen by the s/c. \#rot: number of rotating particles individuated in each dataset. \#im: number of images analysed for each dataset. \#tot: total number of particles in the dataset. \%rot: percentage of rotating particles with respect to the total. $f$: average mode of the rotating frequency. a/b:  average aspect ratio of the particles as defined in Section \ref{flattening}. bkg: value of the  background averaged over each dataset in W/(m$^2$nm sr), computed on an area of 30 pixel radius of each image. Since the illumination condition  is fixed the background variation among each dataset is negligible. The exposure time of each image is 12.5 s}
    
    \label{table:periods}  
\end{table}

 \end{landscape}

\onecolumn
\begin{appendix}

\section{Translational motion of the dust}\label{AppendixA}

\vspace{1cm}
 \includegraphics[width=0.33\textwidth]{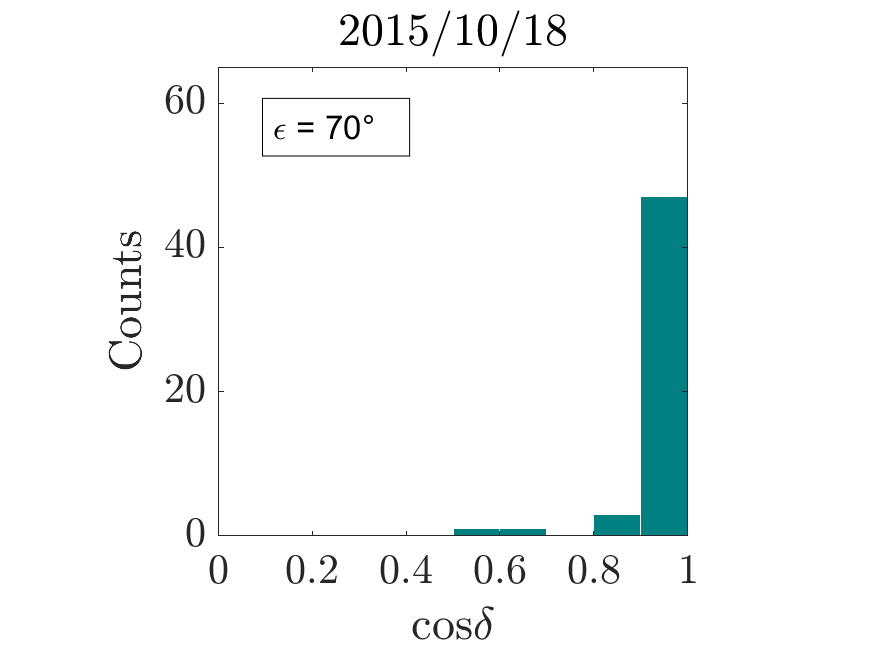}
 \includegraphics[width=0.33\textwidth]{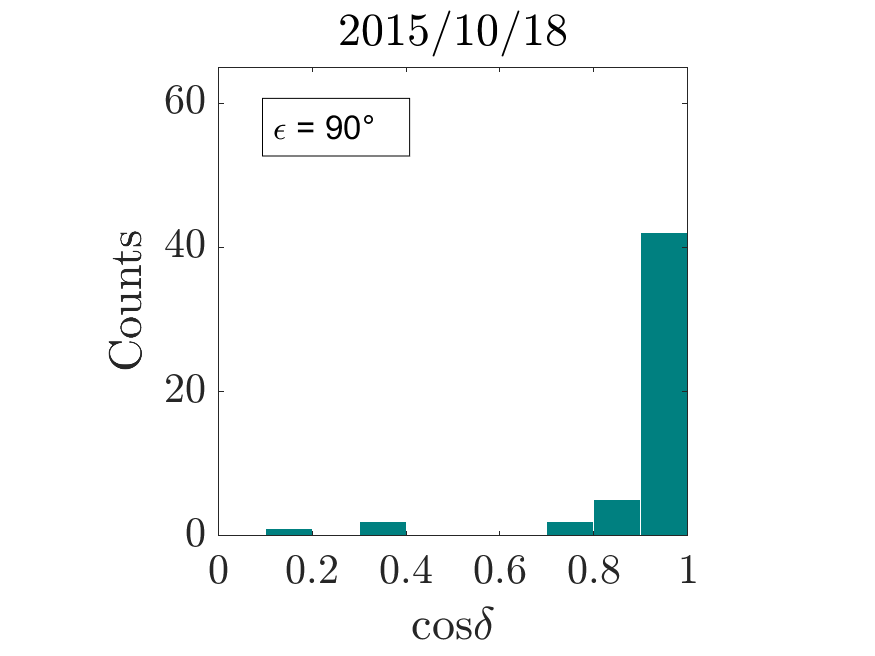}
 \includegraphics[width=0.33\textwidth]{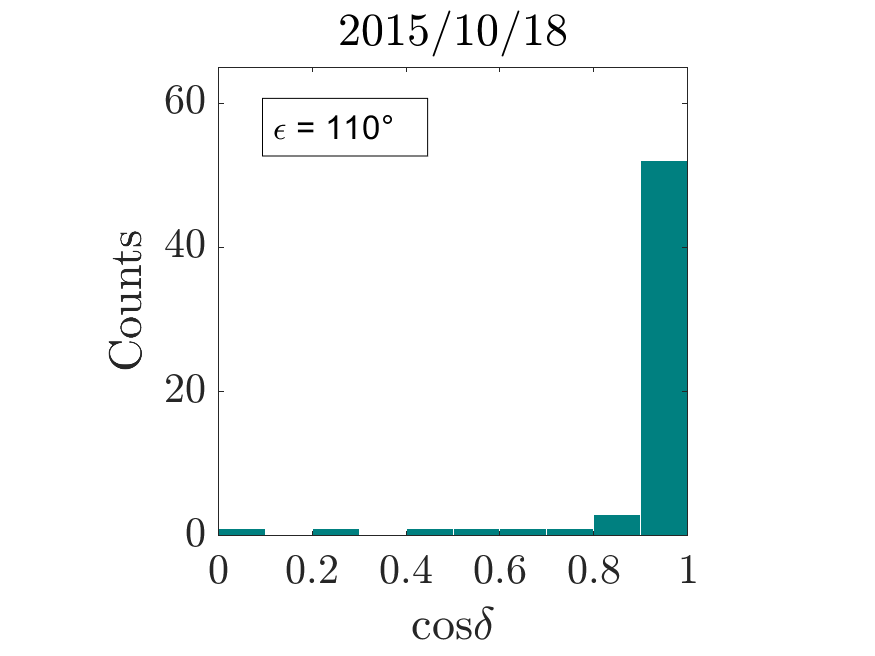}
 \center
 \includegraphics[width=0.33\textwidth]{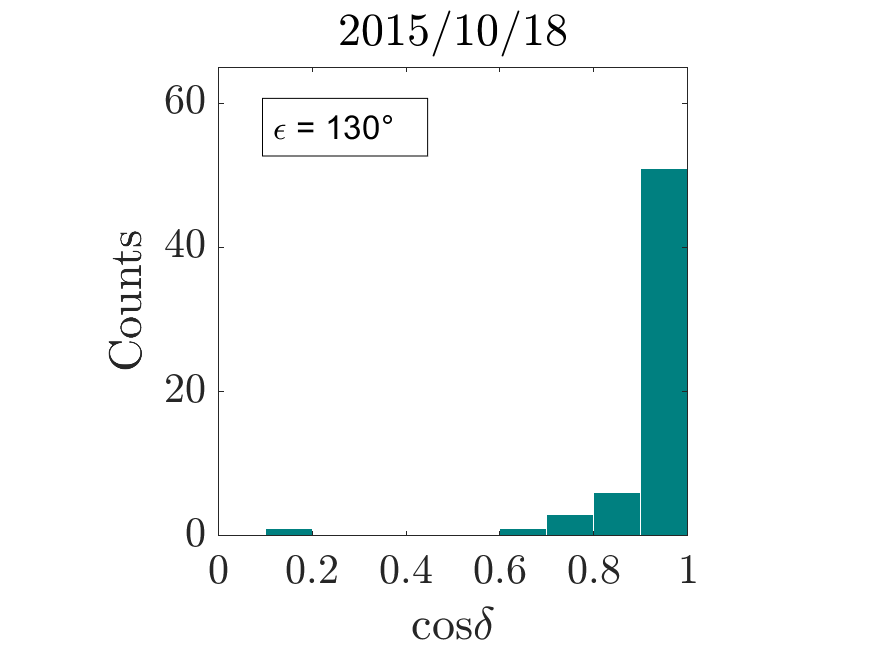}
 \includegraphics[width=0.33\textwidth]{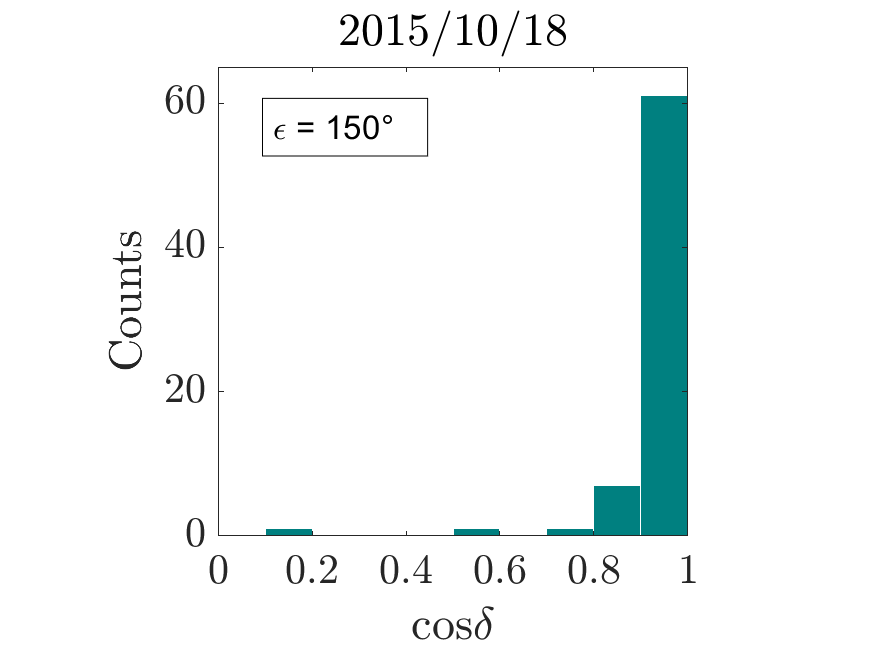}
 \captionof{figure}{The histograms show the value of the cosine of the angle  $\delta$ between  the velocity vector of the particle on the image plane,  and the projected position vector of the nucleus  on the plane of the image, in the NAC reference system, for the set   STP 078 GRAIN\_COLOR\_004.}
\label{fig:radial78}
\vspace{1cm}
\includegraphics[width=0.33\textwidth]{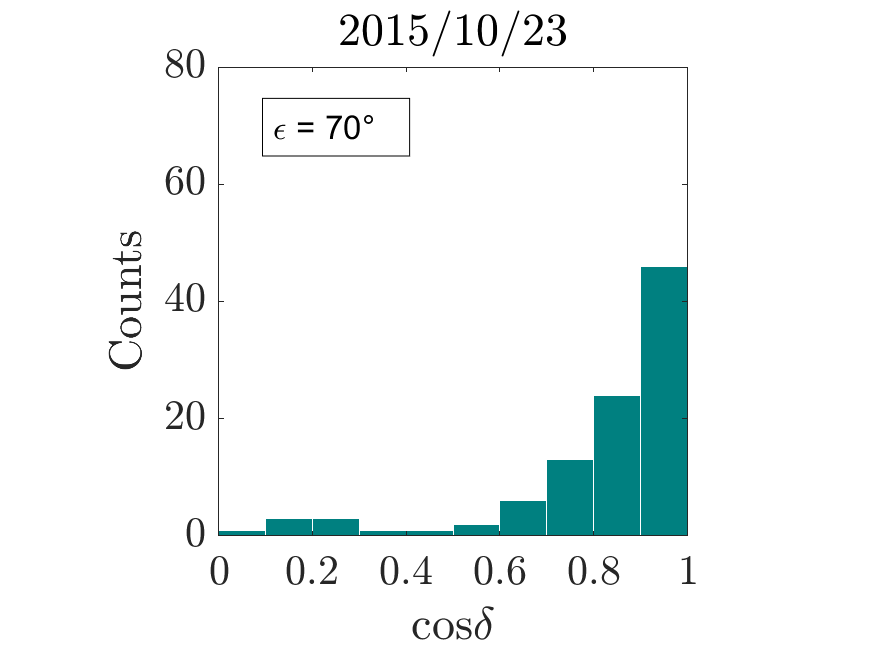}
\includegraphics[width=0.33\textwidth]{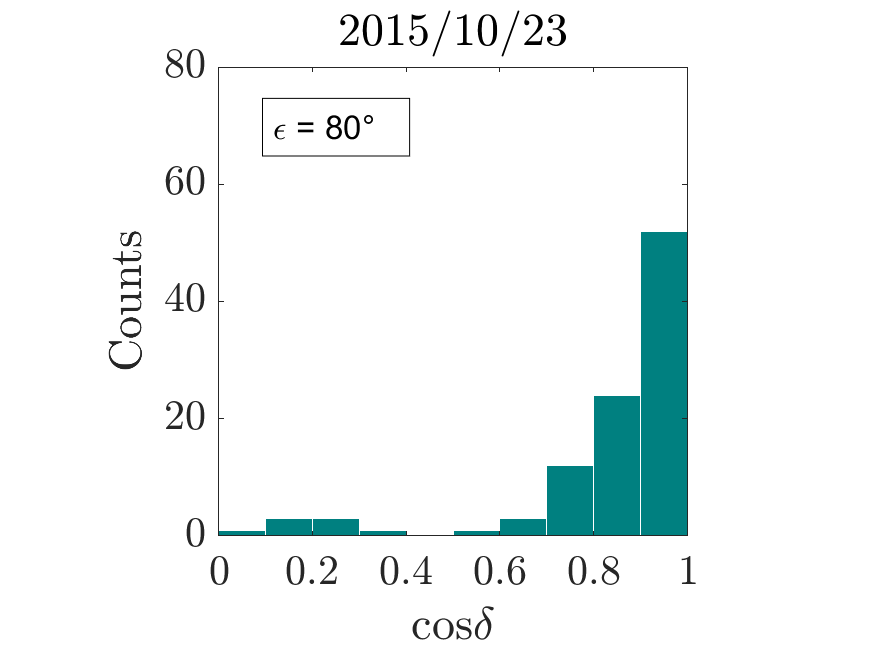}
\includegraphics[width=0.33\textwidth]{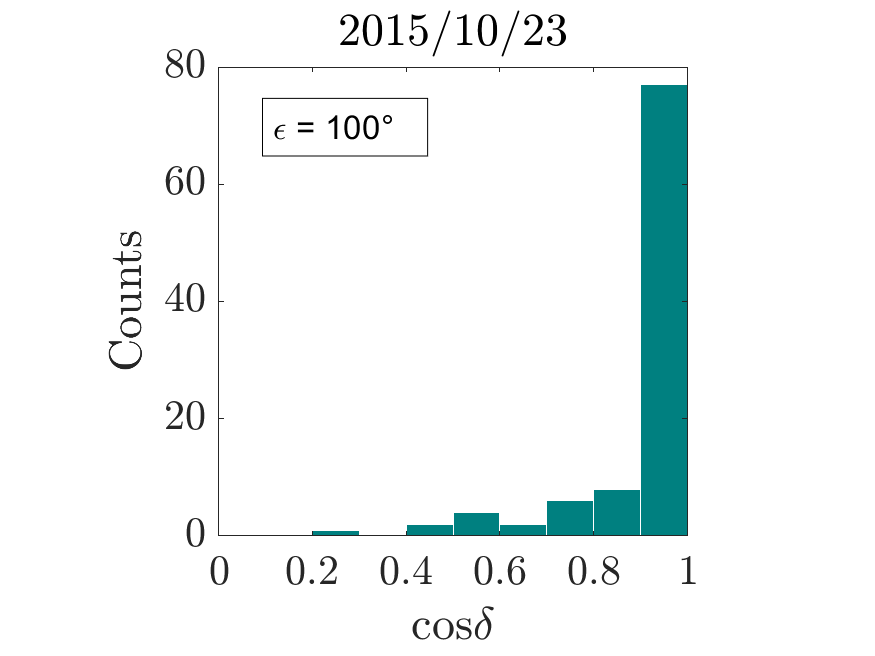}
\center
\includegraphics[width=0.33\textwidth]{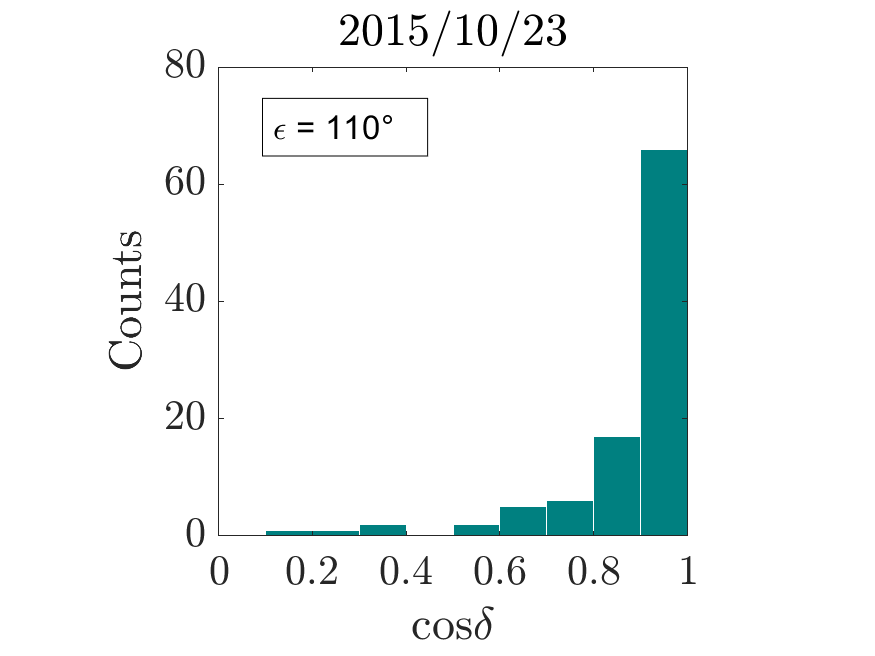}
\includegraphics[width=0.33\textwidth]{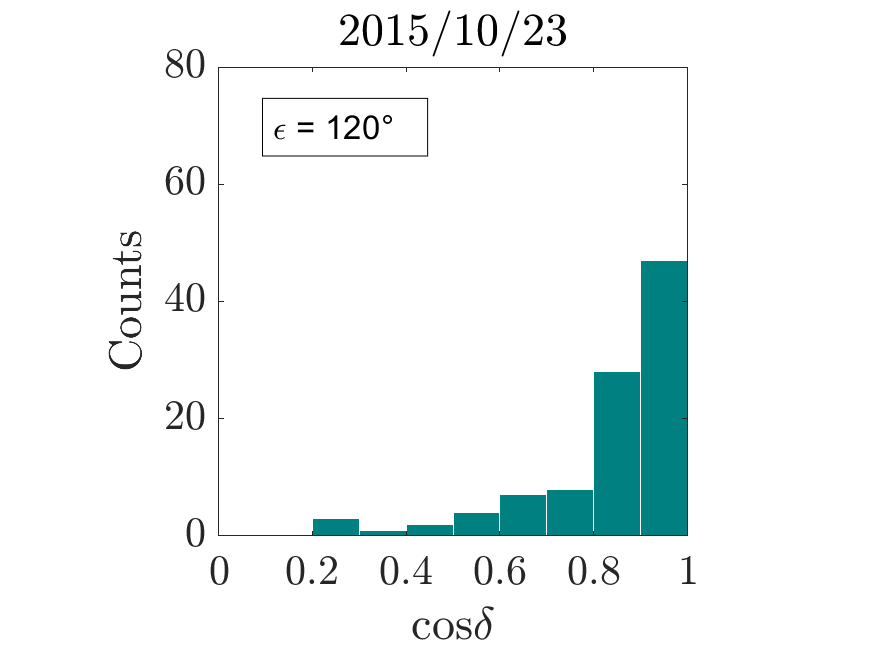}
\includegraphics[width=0.33\textwidth]{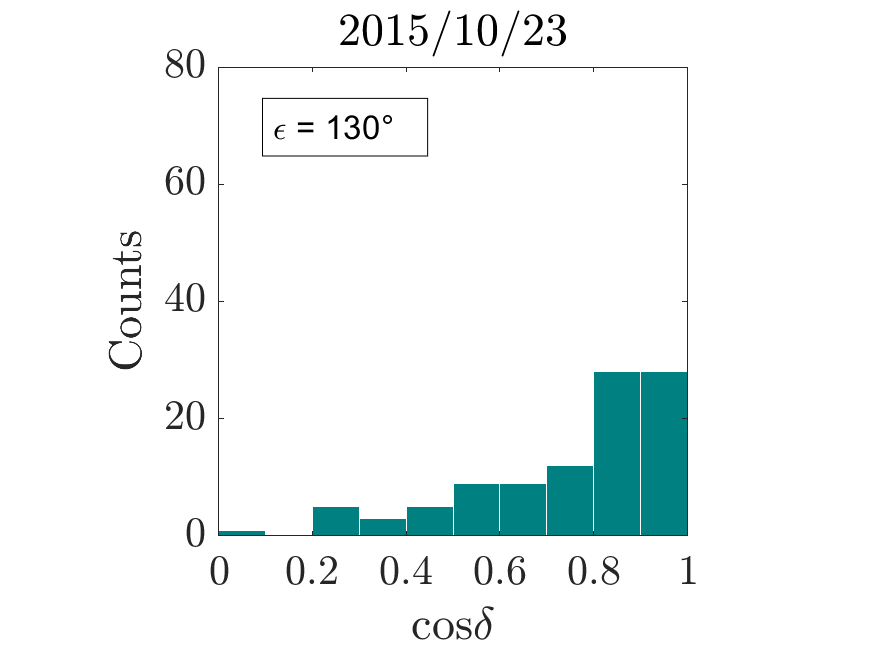}
 \captionof{figure}{The histograms show the value of the cosine of the angle  $\delta$  for the set   STP 079 GRAIN\_COLOR\_001.}     \label{fig:radial79}
\includegraphics[width=0.33\textwidth]{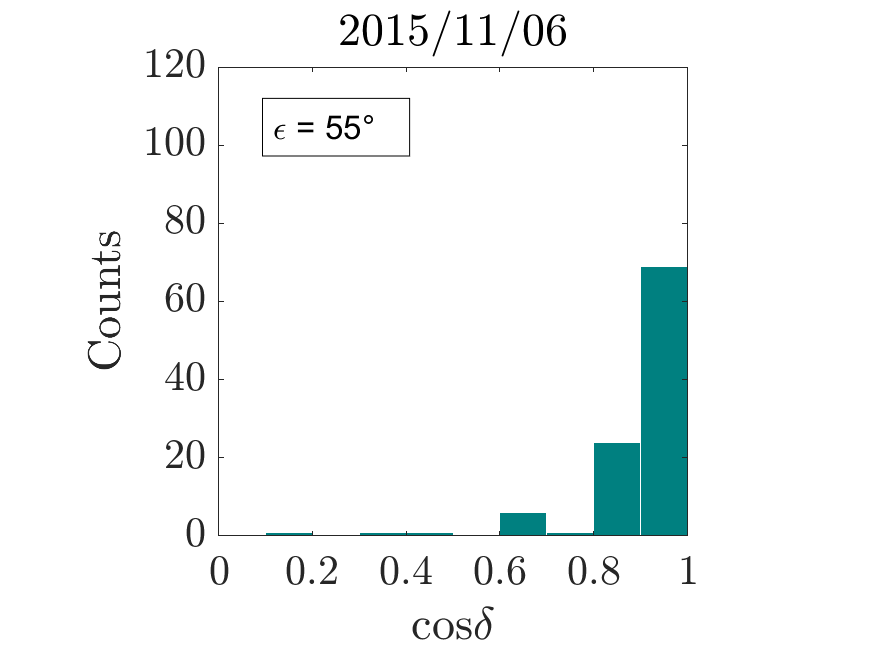}
\includegraphics[width=0.33\textwidth]{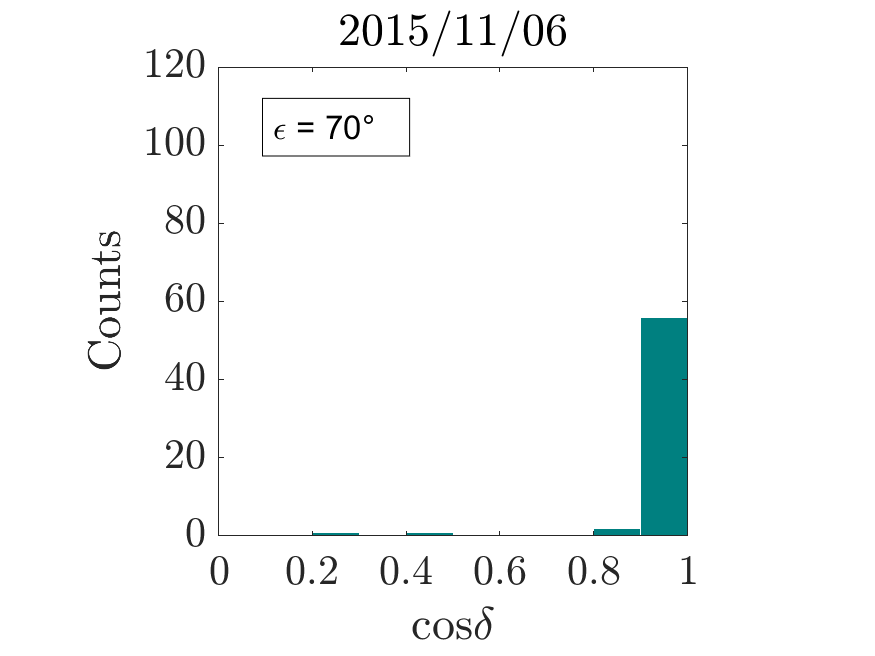}
\includegraphics[width=0.33\textwidth]{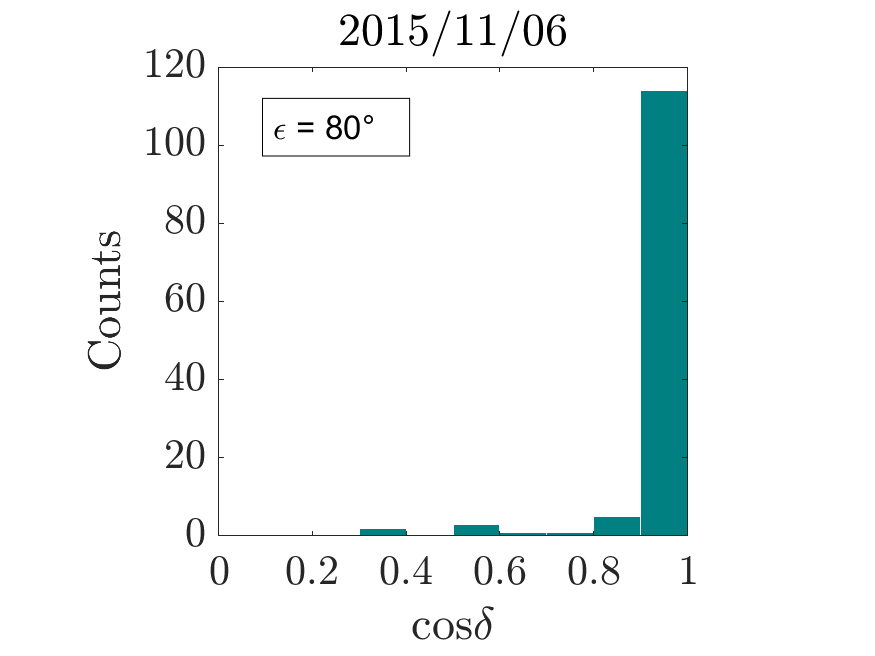}
\center
\includegraphics[width=0.33\textwidth]{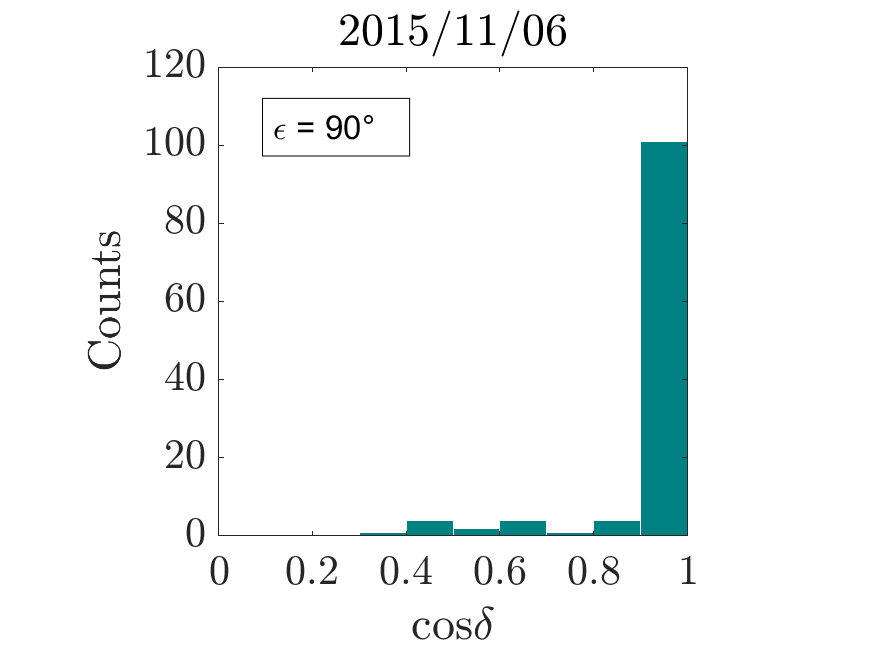}
\includegraphics[width=0.33\textwidth]{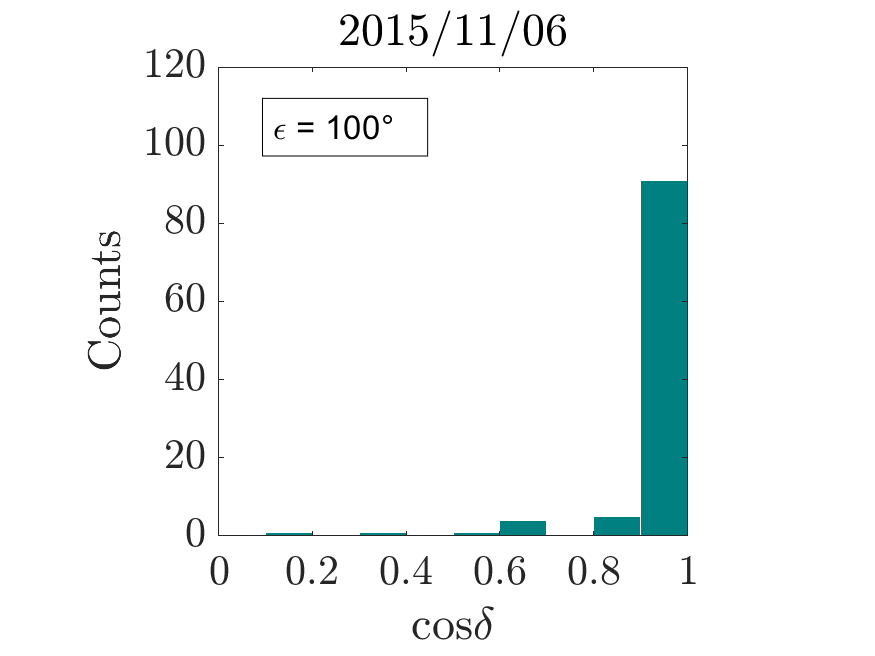}
\includegraphics[width=0.33\textwidth]{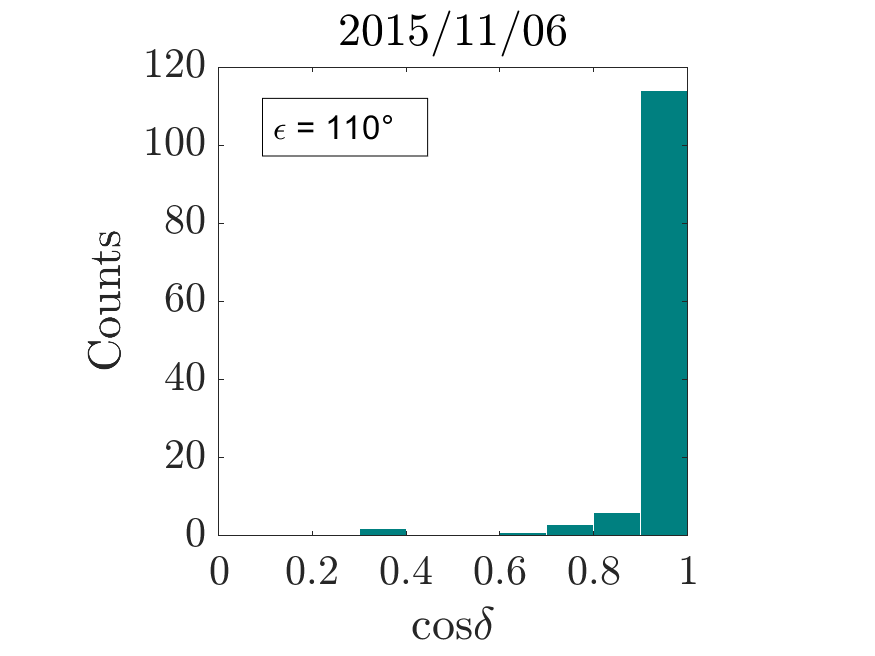}
\center
\includegraphics[width=0.33\textwidth]{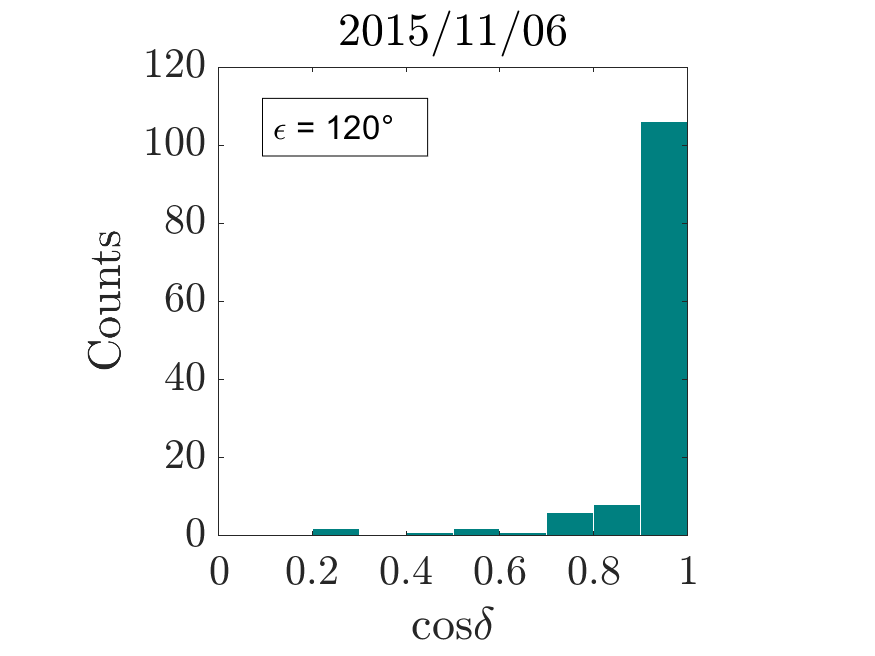}
\includegraphics[width=0.33\textwidth]{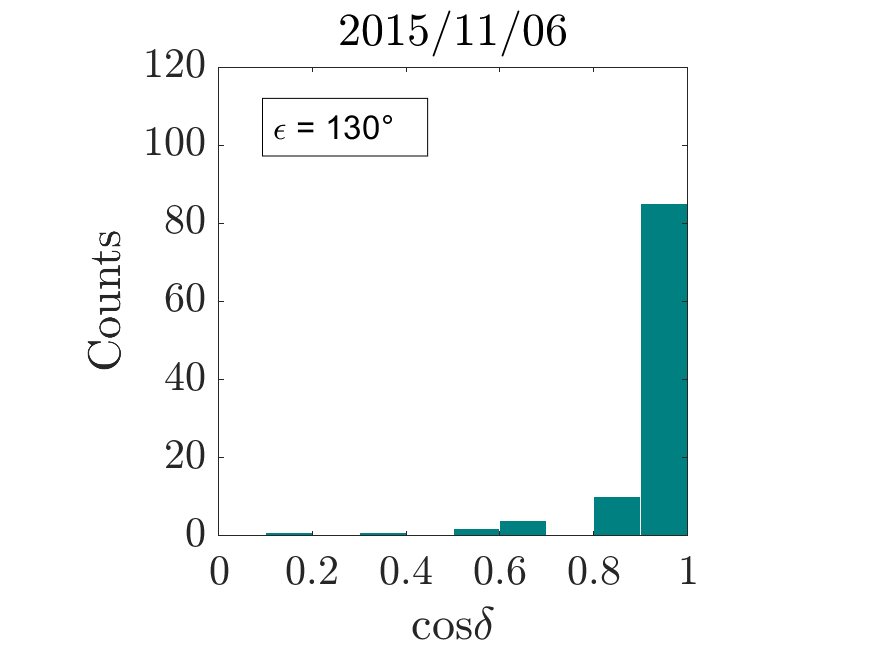}
\includegraphics[width=0.33\textwidth]{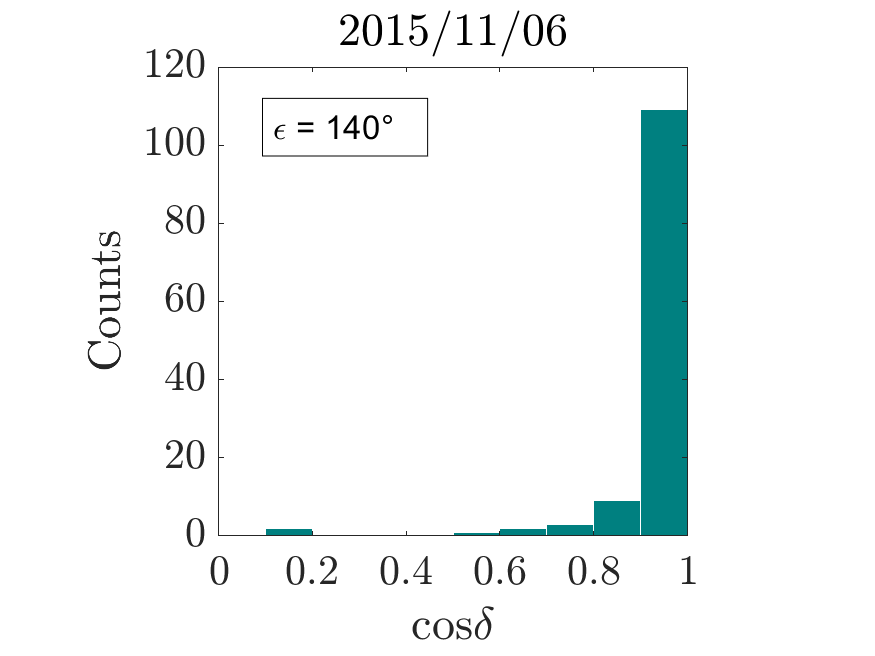}
\captionof{figure}{The histograms show the value of the cosine of the angle  $\delta$   for the set  STP 081 GRAIN\_COLOR\_002.}
\label{fig:radial81}
\includegraphics[width=0.33\textwidth]{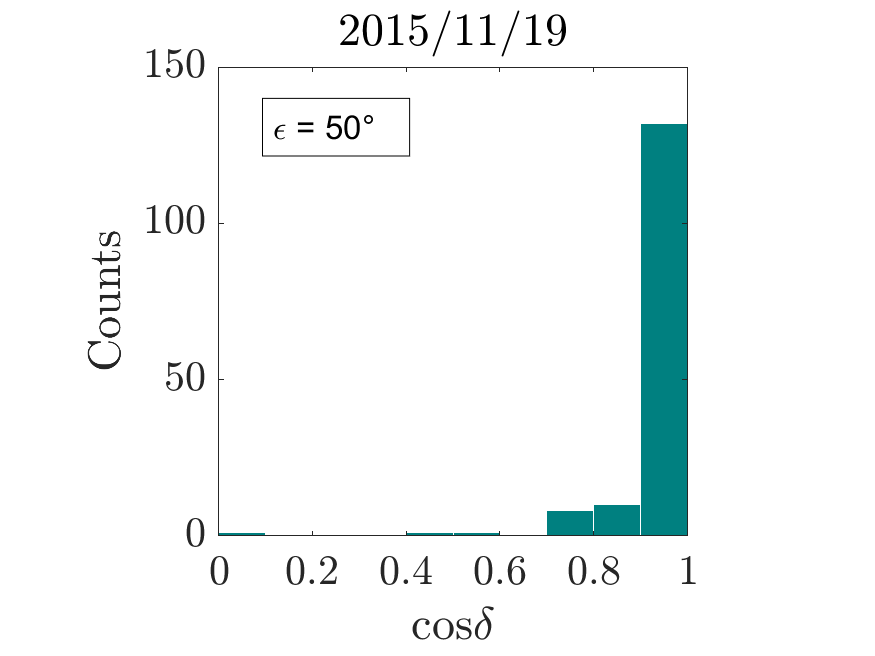}
\includegraphics[width=0.33\textwidth]{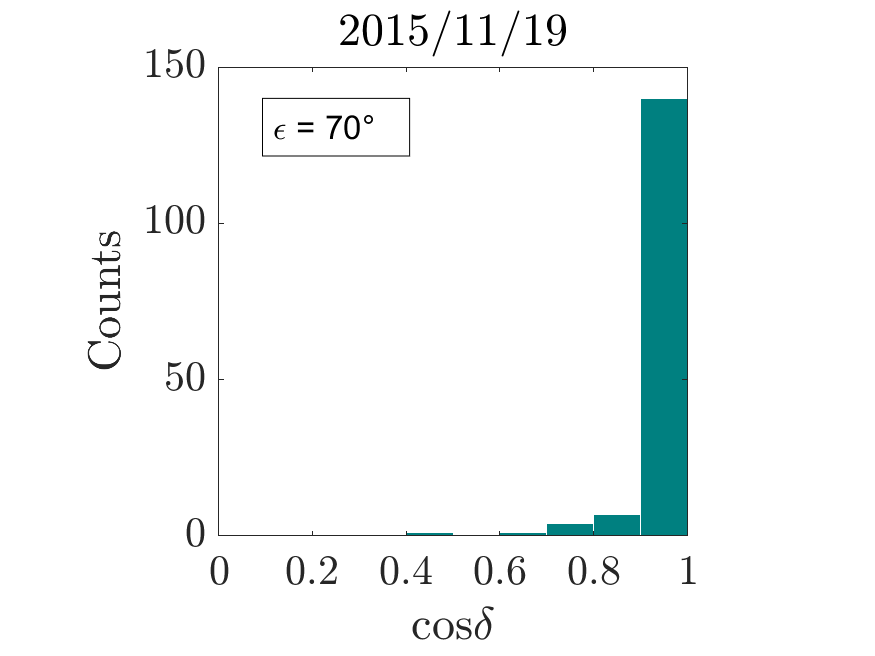} 
\includegraphics[width=0.33\textwidth]{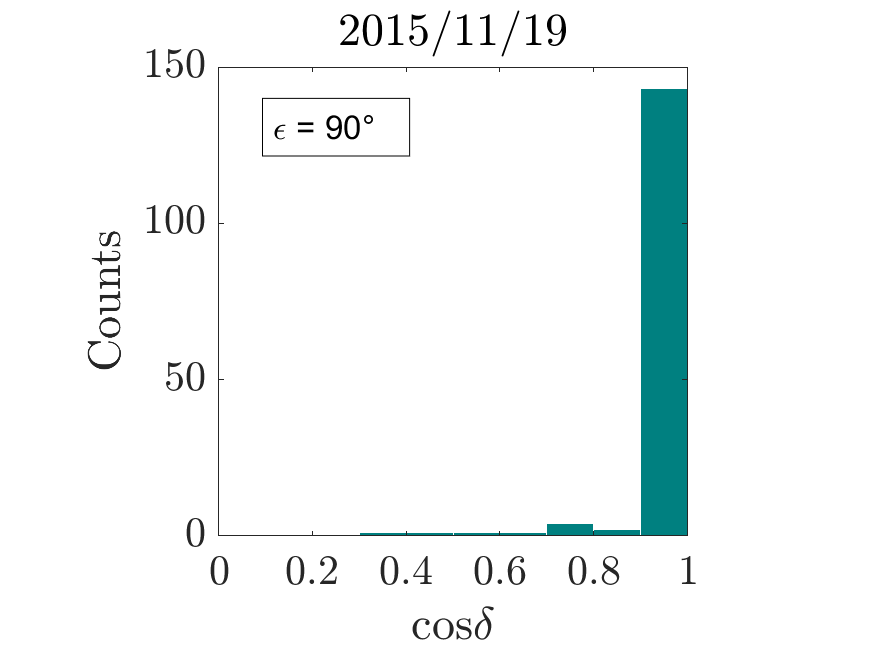}
\center
\includegraphics[width=0.33\textwidth]{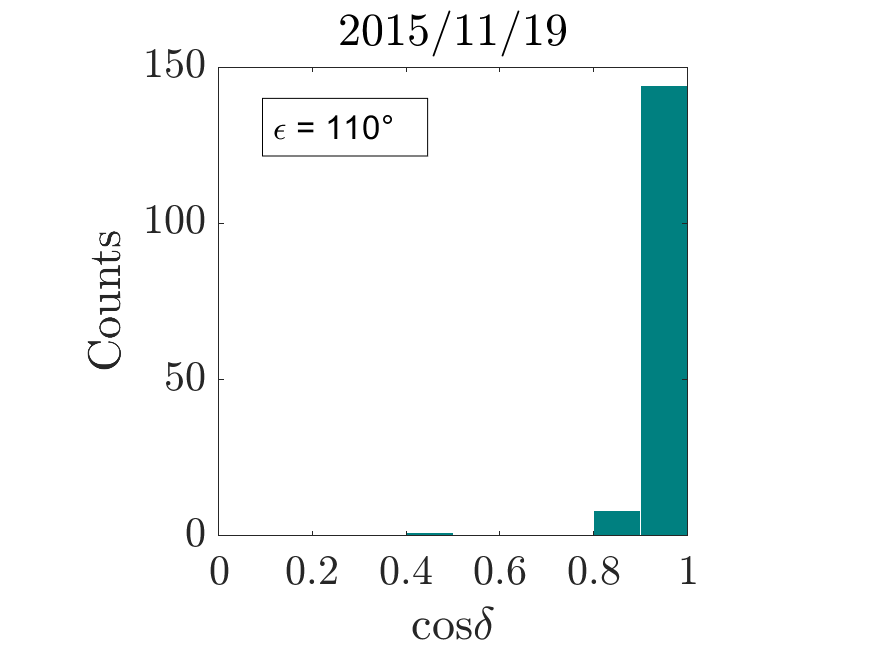}
\includegraphics[width=0.33\textwidth]{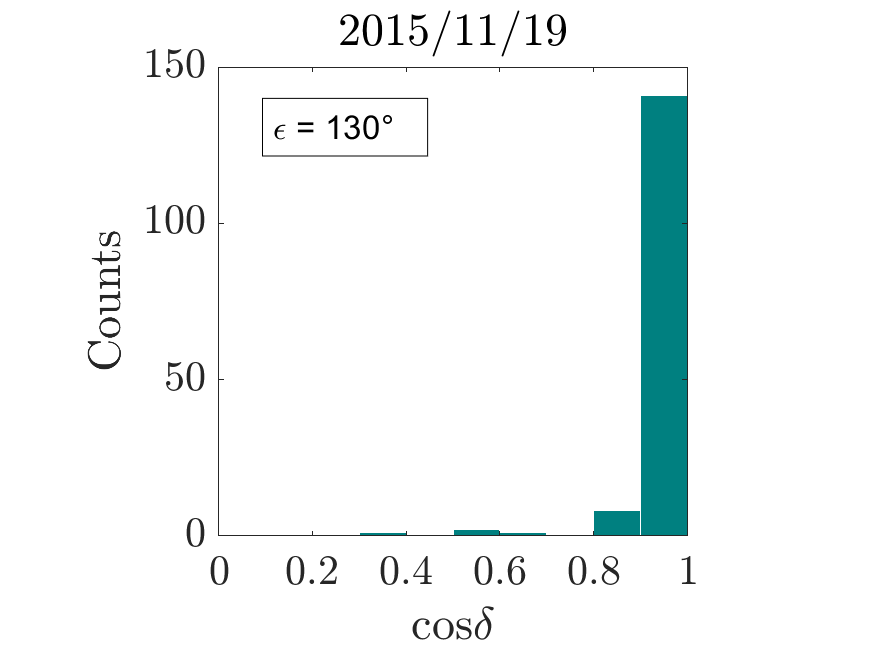}
\captionof{figure}{The histograms show the value of the cosine of the angle  $\delta$ for the set  STP 083 GRAIN\_COLOR\_001.}   
\label{fig:radial83}

\section{Scattering simulations}\label{AppendixB}

    \begin{tabular}{c|c|c|c|c}
    \hline
          $r_h$ (AU) & a (mm) & d$_{s/c}$ (km)  &   Silicates Radiance   &     Organics Radiance \\
        \hline\hline
      \multirow{6}{*}{1.3} &  \multirow{6}{*}{0.1}       & 1 &  $8.2\cdot10^{-7}$ & $3.0\cdot10^{-7}$\\
 &      & 2   & $2.1\cdot10^{-7}$ & $7.5\cdot10^{-8}$ \\   
 &    & 5 & $3.3\cdot10^{-8}$ &  $1.2\cdot10^{-8}$\\
&      & 10  & $8.2\cdot10^{-9}$&$3.0\cdot10^{-9}$  \\
&      & 15  & $3.7\cdot10^{-9}$& $1.3\cdot10^{-9}$  \\
 &      & 20  & $2.1\cdot10^{-9}$& $7.5\cdot10^{-9}$  \\
                    \hline
            \multirow{6}{*}{1.3} &  \multirow{6}{*}{1}       & 1 &  $1.0\cdot10^{-4}$ & $3.0\cdot10^{-5}$\\
 &      & 2   & $2.5\cdot10^{-5}$ & $7.5\cdot10^{-6}$ \\   
 &    & 5 & $4.0\cdot10^{-6}$ &  $1.2\cdot10^{-6}$\\
&      & 10  & $1.0\cdot10^{-6}$&$3.0\cdot10^{-7}$  \\
&      & 15  & $4.4\cdot10^{-7}$& $1.3\cdot10^{-7}$  \\
 &      & 20  & $2.5\cdot10^{-7}$& $7.5\cdot10^{-8}$  \\
                    \hline
      \multirow{6}{*}{1.3} &  \multirow{6}{*}{10}       & 1 &  $7.6\cdot10^{-3}$ & $2.9\cdot10^{-3}$\\
 &      & 2   & $1.9\cdot10^{-3}$ & $7.3\cdot10^{-4}$ \\   
 &    & 5 & $3.0\cdot10^{-4}$ &  $1.2\cdot10^{-4}$\\
&      & 10  & $7.6\cdot10^{-5}$&$2.9\cdot10^{-5}$  \\
&      & 15  & $3.4\cdot10^{-5}$& $1.3\cdot10^{-5}$  \\
 &      & 20  & $1.9\cdot10^{-5}$& $7.3\cdot10^{-6}$  \\
      \hline\hline

                    
      \multirow{6}{*}{1.7} &  \multirow{6}{*}{0.1}       & 1 &  $4.8\cdot10^{-7}$ & $1.8\cdot10^{-7}$\\
 &      & 2   & $1.2\cdot10^{-7}$ & $4.5\cdot10^{-8}$ \\   
 &    & 5 & $1.9\cdot10^{-8}$ &  $7.2\cdot10^{-9}$\\
&      & 10  & $4.8\cdot10^{-9}$&$1.8\cdot10^{-9}$  \\
&      & 15  & $2.1\cdot10^{-9}$& $8.0\cdot10^{-10}$  \\
 &      & 20  & $1.2\cdot10^{-9}$& $4.5\cdot10^{-10}$  \\
                    \hline
            \multirow{6}{*}{1.7} &  \multirow{6}{*}{1}       & 1 &  $5.8\cdot10^{-4}$ & $1.8\cdot10^{-5}$\\
 &      & 2   & $1.5\cdot10^{-5}$ & $4.5\cdot10^{-6}$ \\   
 &    & 5 & $2.3\cdot10^{-6}$ &  $7.2\cdot10^{-7}$\\
&      & 10  & $5.8\cdot10^{-7}$&$1.8\cdot10^{-7}$  \\
&      & 15  & $2.6\cdot10^{-7}$& $8.0\cdot10^{-8}$  \\
 &      & 20  & $1.5\cdot10^{-7}$& $4.5\cdot10^{-8}$  \\
                    \hline
      \multirow{6}{*}{1.7} &  \multirow{6}{*}{10}       & 1 &  $4.5\cdot10^{-3}$ & $1.8\cdot10^{-3}$\\
 &      & 2   & $1.1\cdot10^{-3}$ & $4.0\cdot10^{-4}$ \\   
 &    & 5 & $1.8\cdot10^{-4}$ &  $7.0\cdot10^{-5}$\\
&      & 10  & $4.5\cdot10^{-5}$&$1.8\cdot10^{-5}$  \\
&      & 15  & $2.0\cdot10^{-5}$& $7.8\cdot10^{-6}$  \\
  &      & 20  & $1.1\cdot10^{-5}$& $4.4\cdot10^{-6}$  \\                    \hline\hline
                  \end{tabular}

     \begin{tabular}{c|c|c|c|c}
    \hline
           $r_h$ (AU) & a (mm) & d$_{s/c}$ (km)  &   Silicates Radiance   &     Organics Radiance \\
        \hline\hline
      \multirow{6}{*}{2.0} &  \multirow{6}{*}{0.1}       & 1 &  $3.5\cdot10^{-7}$ & $1.3\cdot10^{-7}$\\
 &      & 2   & $8.8\cdot10^{-8}$ & $3.3\cdot10^{-8}$ \\   
 &    & 5 & $1.4\cdot10^{-9}$ &  $5.2\cdot10^{-9}$\\
&      & 10  & $3.5\cdot10^{-9}$&$1.3\cdot10^{-9}$  \\
&      & 15  & $1.5\cdot10^{-9}$& $5.8\cdot10^{-10}$  \\
 &      & 20  & $8.8\cdot10^{-10}$& $3.3\cdot10^{-10}$  \\
                    \hline
            \multirow{6}{*}{2.0} &  \multirow{6}{*}{1}       & 1 &  $4.2\cdot10^{-5}$ & $1.3\cdot10^{-5}$\\
 &      & 2   & $1.1\cdot10^{-5}$ & $3.3\cdot10^{-6}$ \\   
 &    & 5 & $1.7\cdot10^{-6}$ &  $5.2\cdot10^{-7}$\\
&      & 10  & $4.2\cdot10^{-7}$&$1.3\cdot10^{-7}$  \\
&      & 15  & $1.9\cdot10^{-7}$& $5.8\cdot10^{-8}$  \\
 &      & 20  & $1.1\cdot10^{-8}$& $3.3\cdot10^{-8}$  \\
                    \hline
      \multirow{6}{*}{2.0} &  \multirow{6}{*}{10}       & 1 &  $3.2\cdot10^{-3}$ & $1.3\cdot10^{-3}$\\
 &      & 2   & $8.0\cdot10^{-4}$ & $3.2\cdot10^{-4}$ \\   
 &    & 5 & $1.3\cdot10^{-4}$ &  $5.2\cdot10^{-5}$\\
&      & 10  & $3.2\cdot10^{-5}$&$1.3\cdot10^{-5}$  \\
&      & 15  & $1.4\cdot10^{-5}$& $5.7\cdot10^{-6}$  \\
 &      & 20  & $8.0\cdot10^{-6}$& $3.3\cdot10^{-6}$  \\ 
                    \hline\hline
                \end{tabular}
  \caption{In this table  we report the expected spectral radiance of a dust particle, for different combinations of the scattering code parameters.  $r_h$: the heliocentric distance, a: the size of the particles, $d_{s/c}$: the distance of the particle from the spacecraft. Silicates and Organics Radiance are in units of W/(m$^2$nm sr).  The refractive index of the silicate material is $m = 1.6 + i10^{-5}$, and the refractive index of the organic material is $m = 1.3 + i10^{-2}$.}
    \label{tab:scattering_model}
                    
  \end{appendix}






\end{document}